\documentstyle[aps,prd,preprint,floats,tighten,eqsecnum,epsf]{revtex}

\newcommand{\beq}{\begin{equation}}
\newcommand{\eeq}{\end{equation}}
\newcommand{\beqa}{\begin{eqnarray}}
\newcommand{\eeqa}{\end{eqnarray}}

\begin{document}
\tightenlines

\preprint{\vtop{\hbox{RU99-11-B}\hbox{UM-TH-99-11}\hbox{hep-ph/9912409}
\vskip24pt}}

\title{The Transition Temperature to the Superconducting Phase of QCD at
High Baryon Density}
\author{William E. Brown${}^a$, James T. Liu${}^{b}$
and Hai-cang Ren${}^{a,c}$ \\ [4mm]}
\address{${}^a$ Department of Physics, The Rockefeller University,\\
 1230 York Avenue, New York, NY 10021.\\[2mm]}
\address{${}^b$ Randall Laboratory of Physics, University of Michigan,\\
Ann Arbor, MI 48109.\\ [2mm]}
\address{${}^c$ Department of Natural Science, Baruch College of CUNY,\\
New York, NY 10010.}

\maketitle

\begin{abstract}
Recent interest in the study of color superconductivity has focused on
the regime of high baryon density where perturbative QCD may be
employed.  Based on the dominant one-gluon-exchange interaction, both
the transition temperature and zero temperature gap have been determined
to leading order in the coupling, $g$.  While the leading non-BCS behavior,
$T_C\sim\mu g^{-5}e^{-\kappa/g}$, is easily obtained, the pre-exponential
factor has proved more difficult to evaluate.  Focusing on the transition
temperature, we present a perturbative derivation of this factor, exact
to leading order in $g$.  This approach is first motivated by the study
of a toy model and involves working to second order in the perturbative
expansion.  We compare this result to the zero temperature gap.
Additionally, we extend the analysis to the case of higher angular
momentum for longitudinal and transverse quark pairing.
\end{abstract}


\newpage

\widetext

\section{Introduction.}
Previous studies of the quark-gluon plasma at high baryon density
have indicated the possibility of a pairing instability which may induce a
novel superconducting phase of QCD \cite{Barrois,Frau,bailin1984}.  Based on
conventional superconductivity theory, it is known that even the weakest
attractive interaction between quarks would be sufficient to form a
superconducting pair; for $SU(3)_c$ QCD this is provided by the
${\bf \bar 3}$ channel in the diquark interaction.  Motivated by studies
of QCD under unusual conditions, interest in this phenomenon has recently
been revived \cite{alford1998a,rapp1998}.  In fact the phase structure of
matter in this regime has important physical consequences for the behaviour
of dense astrophysical objects as well as for heavy ion collisions and has
raised fundamental questions about the nature of QCD.

In general, for $SU(N)_c$ QCD with $N_f$ light flavors, the combined
symmetry of the theory is initially $SU(N)_c \times SU(N_f)_L \times
SU(N_f)_R \times U(1)_B$.  However, a diquark condensate,
$\langle q q\rangle$, would necessarily carry both color and flavor
charge, and hence break the above symmetry.  The pattern of symmetry
breaking is especially interesting for the case of three flavours and
colors, in which case the initial
$SU(3)_c \times SU(3)_L \times SU(3)_R \times U(1)_B$ symmetry is broken to
the global diagonal subgroup $SU(3)_{c+L+R}$
\cite{alford1998b,schafer1998,ABR99a,SWa,EHHS}.  Since the condensate
remains symmetric under combined color and flavor rotations, it ``locks''
the color and flavor degrees of freedom.  Further interest in the
color-flavor locked state has arisen from the observation that many
properties such as quantum numbers and roughly similar masses are
shared by the bosons produced by this symmetry breaking and hadronic
matter, suggesting a connection between these two phases of QCD
\cite{schafer1998}.  These results have led to further investigation of the
color-flavor locked phase \cite{ABR99b} and the development of an
effective field theory for the Nambu-Goldstone bosons produced by the
symmetry breaking \cite{HRZ,CG,RWZ}.

The nature of the phase transition to the quark gluon plasma and the
hadronic phases of QCD has also been studied.  Due to competition
between confinement and chiral condensation on one side and the
formation of a diquark condensate on the other, it is expected that
the transition to the hadronic phase is of first order
\cite{alford1998a,BR99,diak,pisarski1998,Raja}.  Furthermore the phase
transition to the quark gluon plasma has also been identified as first order
\cite{BR99,Raja}.  Other physical implications that have been studied
include the effect of differing densities of $s$ compared to $u$ and $d$
quarks, \cite{bedaque}, and the effect of a color-superconducting core
upon the magnetic field of neutron stars, observed to be stable over
very long time scales \cite{ABR99c,BSS,BKV,Mats}.

Since the study of color superconductivity involves working at non-zero
chemical potential, it has not been amenable to numerical simulations on a
lattice.  Instead, a number of analytic methods have been developed to
qualitatively probe the color superconducting state
\cite{alford1998a,rapp1998,alford1998b,ABR99a,BR99,diak,RSSV,EHSa,EHSb,SWb}.
All of these approaches, to some extent, have difficulties with possible
Ansatz dependence, gauge invariance, the Meissner effect as well as the
non-linearity of the gap equation, thus making calculations of the
condensate properties or the transition temperature somewhat difficult.
The consensus of all these models, however, is that, as expected, the
strength of the quark-quark interaction in the most attractive channel
is of fundamental importance in determining the size of the condensate
and hence the properties of the superconducting state \cite{Raja}.

Quantitative results on color superconductivity may be obtained in the
limit of ultra-high baryon density, where perturbative QCD is expected
to be valid.  In this regime, the dominant attractive interaction arises
by single gluon exchange and may be incorporated in a Schwinger-Dyson
analysis to yield either superconducting gap equations or conditions on
the transition temperature.  Because of the long range nature of the
single gluon exchange interaction, the coupling dependence of the gap
$\Delta$ takes on the non-BCS form \cite{son1998},
\begin{equation}
\Delta \sim c{\mu \over g^5}e^{-\kappa/g},
\end{equation}
and this behavior holds for the transition temperature $T_C$ as well.
In this paper we refine the analysis of the transition temperature and
present an exact calculation of $T_C$ to leading order in the coupling
of the exponential ($\kappa$) and pre-exponential ($c$) factors.  This
is carried out for both longitudinal (LL and RR) and transverse (LR)
quark pairings of arbitrary angular momentum, $J$.  We find that the
$s$-wave $(J=0)$ channel, in which only longitudinal pairings of quarks
may occur, is dominant.  For $SU(N)_c$ QCD the general result takes
the form,
\begin{equation}
\label{TC}
\pi k_BT_C^J={\mathcal C}_1^{\vphantom{J}}{\mathcal C}_2^J 256 \pi^4
\left( \frac{2}{N_f} \right)^{\frac{5}{2}}
\frac{\mu}{g^5}
e^{-\sqrt{6N\over N+1}{\pi^2\over g}},
\end{equation}
where $N_f$ is the number of flavors, $\mu$ is the
chemical potential and $g$ is the running coupling constant evaluated at
$\mu$.  The largest factor,
\beq
\label{sonsfactor}
\Delta_0 = 256 \pi^4 \left( \frac{2}{N_f} \right)^{\frac{5}{2}}
\frac{\mu}{g^5}
e^{-\sqrt{6N\over N+1}{\pi^2\over g}},
\eeq
was first calculated in \cite{son1998,SWc,pisarski1999b}, initially
using an elegant
renormalisation group analysis.  The pre-factors ${\cal C}_1^{\vphantom{J}}$
and ${\cal C}_2^J$ represent angular momentum channel independent and
dependent corrections respectively, the calculation of which are the focus
of this paper.
This novel non-BCS dependence of $\Delta_0$ on the coupling has
been confirmed in \cite{BLR99,hs,pisarski1999c}.
In addition, the leading order ratio of the transition temperature to the
zero temperature gap (in the $s$-wave channel) was investigated in
\cite{pisarski1999c} and found to be identical to the BCS case,
namely $k_BT = {e^\gamma\over\pi}\Delta$ where
$\gamma=0.5772\ldots$ is the Euler constant.

In order to exactly determine the pre-factors ${\mathcal C}_1^{\vphantom{J}}$
and ${\mathcal C}_2^J$, which are of ${\mathcal O}(1)$, it is most
straightforward to approach the transition temperature from the normal
phase rather than from within the superphase.  In the normal phase
there is no condensate, no Meissner effect, the problem is linear and
there is an established framework to ensure gauge invariance.
Following the formulation developed in \cite{gorkov1961} for the
non-relativistic Fermi gas, the Dyson-Schwinger equation for the
scattering of two quarks can be cast into a general eigenvalue
problem.  This is outlined in \cite{BLR99} and more details are given
in this paper.  The general eigenvalue problem can be solved to give a
set of eigenvalues, $\{\lambda_j^2\}$, in terms of the parameters of the
theory.  The transition to the superphase is then characterised by the
locus on the phase diagram where the smallest eigenvalue becomes one.
Physically, this condition defines the pole in the solution to the
integral equation which represents the formation of a bound state of
quarks.

The pre-factor ${\mathcal C}_1$ is found to be of ${\mathcal O}(1)$, and has
the form:
\begin{eqnarray}
{\mathcal C}_1 &=& 2e^\gamma \exp\left[{-\frac{1}{16}(\pi^2 + 4)(N-1)}\right]
\nonumber\\
&\simeq&0.629\qquad\hbox{for $N=3$}.
\end{eqnarray}
To understand the physical origins of this factor, it can be decomposed
into three parts, ${\mathcal C}_1 = c_1' c_1'' c_1'''$, each of which
are entirely due to single gluon exchange.  Higher order diagrams have
been shown not to contribute to the leading order of the exponential
or the pre-factor in \cite{BLR99}.  The first contribution, $c_1'$,
originates in the radiative corrections to the quark propagator due to
its self-energy \cite{BLR99},
\beq
\label{c1}
c_1' = \exp \left[{-\frac{1}{16} (\pi^2 + 4) (N-1)}\right].
\eeq
The remaining two contributions, $c_1''$ and $c_1'''$, are
related to the next-to-leading order logarithmic infrared behaviour
overlooked in the obtention of (\ref{sonsfactor}).  The factor $c_1''$
accounts for the sub-leading behaviour lost in the transition from
the Matsubara sum at finite temperature to a continuous energy integral,
\beq
\label{eq:c1pp}
c_1'' = 2e^\gamma.
\eeq
The final factor, $c_1'''$, represents the contribution of all other
approximations made in the obtention of (\ref{sonsfactor}).  Perhaps
somewhat surprisingly, we find no such contributions to leading order
in the coupling, so that in fact
\beq
c_1''' = 1.
\eeq

The second pre-factor, ${\cal C}_2^J$, classifies the dependence of
the critical temperature upon the angular momentum of the pairing.
The dependence upon angular momentum is different for longitudinal and
transverse condensates.  In the dominant $s$-wave channel, for longitudinal
pairs we find ${\mathcal C}_2^{J=0} =1$ whereas, due to conservation of
angular momentum, for transverse pairs we find ${\mathcal C}_2^{J=0}=0$.
Both longitudinal and transverse pairs may exist for higher angular
momentum and the results are summarised below,
\begin{equation}
{\mathcal C}_2^J =
\cases{
 \exp\left[ 3 c_J \right]  & \mbox{longitudinal pairs,  $ J \geq 0$},\cr
\noalign{\vskip2mm}
\exp \left[ \frac{3}{2} \left( c_J + \frac{J}{2 J + 1}
c_{J+1} + \frac{J+1}{2 J +1} c_{J-1} \right) \right]  &
\mbox{transverse pairs, $J \geq 1$,}\cr}
\end{equation}
where,
\beq
c_J =
\cases{
 0  & \mbox{ for  $ J = 0$},\cr
-2 \sum_{n=1}^J \frac{1}{n} &  \mbox{ for $J \neq 0$.}\cr}
\eeq
This indicates the suppression of the higher angular-momentum channels, so
that $s$-wave pairing is dominant and $p$-wave pairing is already down by
about three orders of magnitude.

Before proceeding to the complete derivation of (\ref{TC}), we shall
first illustrate the perturbative formalism in section II and apply it
to an exactly soluble toy model.  Then in section III, the
Dyson-Schwinger equation for the interaction between a longitudinal
pair of quarks of zero angular momentum is studied in the Coulomb
gauge.  The resulting Fredholm equation is cast into a general
eigenvalue problem.  A perturbative solution for this eigenvalue
problem is found in section IV.  Systematic calculation to second
order in this perturbation series captures all of the double and
linear logarithmic infrared behaviour of the eigenvalue that
determines (\ref{TC}) to leading order in the coupling.  The
generalization to non-zero angular momentum and transverse pairings is
made in section V and in section VI we discuss numerical verification
and the gauge invariance of our results.  Concluding remarks are made
in section VII.
%

\section{A Toy Model: The Two-dimensional Cylindrical Well.}
In this section we shall develop the perturbative formalism that will
be used in the following sections.  As shown in \cite{BLR99}, the
Schwinger-Dyson equation determining the transition temperature has the form
of a Fredholm integral equation and may be re-cast as an eigenvalue
problem.  We thus seek a systematic perturbative method of developing the
eigenvalues of the Fredholm equation.  To illustrate this method, we begin by
considering the Schrodinger problem for a two-dimensional cylindrical
well of depth $\kappa^2/a^2$ and radius $a$:
\beq
\label{1}
\left(\frac{\partial^2}{\partial x_i^2} + \frac{\kappa^2}{a^2}
\theta(a - |\vec x|)\right)
\psi(\vec{x}) = \Delta^2 \psi(\vec{x})
\eeq
(where we have normalised the mass so that $2 m = 1$).  As a toy model, it
is easily solved in terms of Bessel functions and matching conditions at the
well boundary.  On the other hand, it may also be converted into an
integral equation of Fredholm type.  The perturbative solution to the
resulting Fredholm equation will serve the basis for generalization to the
real problem of determining the QCD transition temperature.

Note that although two parameters appear in (\ref{1}), namely $a$ and
$\kappa$, only the dimensionless coupling $\kappa$ is physical, as $a$ may
be scaled away by $\vec x\to a\vec x$.  In analogy with the QCD problem,
$\kappa$ represents the QCD coupling $g$, and $a$ represents the inverse
of the chemical potential $\mu$.  The eigenvalue $\Delta^2$ plays the
r\^ole of either the critical temperature or the mass gap of the color
superconductor and in the weak coupling limit has the form $\Delta \propto
(1/a)\exp(-1/\kappa^2)$.  While this behavior is BCS-like
(as may be deduced from
the short-range nature of the cylindrical well potential) some mathematical
features of this toy model nevertheless extend to the long-range QCD case.

In the following, we show that this leading order in $\kappa$ behaviour
of $\Delta$ can be recovered from a perturbative expansion of the
resulting Fredholm kernel, where the zeroth order operator is defined
to capture the leading logarithmic behaviour.  In this case we find that
the exponential and its pre-factor can be determined to leading order in
$\kappa$ by second order perturbation theory.  Furthermore, the exponential
is determined by the leading logarithmic behaviour and the pre-factor is
determined by the sub-leading behaviour uncovered by perturbation theory.

For comparison with the perturbative method, we now
briefly give the exact solution of the model.  From the rotational
symmetry of the problem, only the $s$-wave equation is important.  In
this case the radial Schrodinger equation becomes
\beq
\label{2}
\frac{1}{r} \frac{d}{dr}\left(r\frac{d\psi(r)}{dr}\right)
+ \frac{\kappa^2}{a^2} \theta(a-r) \psi(r) = \Delta^2 \psi(r).
\eeq
The solutions of (\ref{2}) are Bessel functions, both inside and outside
the well.  The eigenvalue problem then results from the matching conditions
at $r=a$.  For a bound state, $\Delta^2 >0$, the matching conditions read,
\beq
\label{2.3}
\sqrt{\kappa^2 - \Delta^2 a^2} \frac{J_1\left(\sqrt{\kappa^2 - \Delta^2 a^2}
\right)}{J_0\left(\sqrt{\kappa^2 - \Delta^2 a^2}  \right)} = \Delta a
\frac{K_1\left(\Delta a \right)}{K_0\left(\Delta a \right)}.
\eeq
To draw out the similarities with superconductivity, we find the
leading order solution in the weak coupling limit, $\Delta a \ll 1$ and
$ \kappa \ll 1 $, to be
\beq
\label{5}
\Delta \simeq \Delta_0 \exp\left(-\frac{2}{\kappa^2}\right),
\eeq
where $\Delta_0 = (2/a) \exp(- \gamma + 1/4)$.

To make contact with the Fredholm problem, we now re-write the toy model
in terms of an integral equation in momentum space.  Using the Fourier
transform of the potential,
\beq
V(\vec{q}\,) = -\frac{\kappa^2}{a^2}\int d^2\vec{x}\,e^{-i\vec{q}\cdot\vec{x}}
\theta(a - |x|) = -2\pi \kappa^2 \frac{J_1(qa)}{qa} ,
\eeq
and wavefunction,
\beq
\psi(\vec{x}\,) = \int \frac{d^2\vec{p}}{(2\pi)^2} e^{i\vec{p}\cdot\vec{x}}
\chi(\vec{p}\,),
\eeq
we get a homogeneous Fredholm equation similar to that
obtained from the Schwinger-Dyson approach to QCD at high density,
\beq
\label{6}
\chi(\vec{p}\,) = -\frac{1}{p^2 + \Delta^2} \int \frac{d^2\vec{q}}{(2\pi)^2}
V(\vec{p} - \vec{q}\,) \chi(\vec{q}\,).
\eeq
Partial wave analysis simplifies the solution
of this integral equation.  Expanding the potential in the
standard way,
\beqa
V(\vec{p} - \vec{q}\,) &=& \sum_m V_m (p,q) e^{im\phi}, \\ \nonumber
V_m(p,q) &=& \frac{1}{2\pi} \int_0^{2\pi} d\phi\, e^{-im\phi}
V(\vec{p} - \vec{q}\,) = - \frac{\kappa^2}{a^2} 2\pi \int_0^a dr\, r
J_m(pr)J_m(qr) ,
\eeqa
we note, for a cylindrically symmetric wavefunction $\chi(p)$, that
only $V_0$ survives the integration over the azimuthal angle in
(\ref{6}) and we have
\beq
\chi(p) = - \frac{1}{2\pi(p^2 + \Delta^2)} \int_0^\infty
dq\,q V_0(p,q) \chi(q).
\eeq

We now consider the general eigenvalue problem,
\beq
\chi(\vec{p}\,) = - \frac{\lambda^2}{p^2 + \Delta^2} \int
\frac{d^2 \vec{q}}{(2\pi)^2} V(\vec{p}-\vec{q}\,) \chi(\vec{q}\,),
\eeq
for which the condition $\lambda^2 = 1$ identifies the specific
problem of (\ref{6}).  We shall find solutions, in the weak coupling
limit, for the eigenvalues $\lambda^2$ in terms of the parameters of
the theory; $\kappa$, $a$ and $\Delta$.
%

Perturbatively it is convenient to work with a real and symmetric
kernel.  Noting that $V_0(p,q)$ is already symmetric in $p$ and $q$ and
that the kernel is real, symmetrization is achieved with the definition,
\beq
f(p) = \sqrt{p (p^2 + \Delta^2)} \chi(p),
\eeq
so that the problem is reduced to the solution of the integral equation,
\begin{equation}
\label{10}
f(p) = \lambda^2 \int_0^{\infty} dq\, {\mathcal K}_S(p,q) f(q),
\end{equation}
where
\begin{equation}
\label{eq:skern}
{\mathcal K}_S(p,q) = -\frac{1}{2\pi}\sqrt{\frac{pq}{(p^2 + \Delta^2)
(q^2 + \Delta^2)}} V_0(p,q)
\end{equation}
is the symmetrized kernel.

We observe that the most divergent region of ${\mathcal K}_S$ is at small
$p$ and $q$.  Since $V_0$ is well behaved at small $p$ and $q$ the dangerous
terms arise in the denominator of (\ref{eq:skern}) and are damped by $\Delta$.
Thus the important contributions to the integral equation arise in the
region $\{p,q\} \approx \Delta$.  To highlight this region, we split the
integral equation into two parts by introducing a scale $\delta$ such
that $\Delta \ll \delta \ll 1/a$.  We first look at the region
$p<\delta$, $q<\delta$ that contains the leading order behaviour.  In this
region, $V_0(p,q) \simeq -\kappa^2 \pi$, so the kernel may be split into
a zeroth order term and correction:
\beq
{\mathcal K}_S(p,q) = {\mathcal K}_S^{\circ}(p,q) + {\mathcal K}_S'(p,q),
\eeq
where
\beq
\label{cut}
{\mathcal K}_S^{\circ}(p,q) = \frac{\kappa^2}{2}
\sqrt{\frac{pq}{(p^2 + \Delta^2)(q^2 + \Delta^2)}}
\theta(\delta - p) \theta(\delta - q)
\eeq
and
\beq
\label{21}
{\mathcal K}_S'(p,q) = \frac{\kappa^2}{a^2} \sqrt{\frac{pq}{(p^2 + \Delta^2)
(q^2 + \Delta^2)}} \int_0^a dr\, r \left[J_0 (pr) J_0 (qr) -
\theta(\delta - p) \theta(\delta - q) \right].
\eeq

To the zeroth order, the integral equation (\ref{10}) reduces to
\beq
f^{\circ}(p) =  {\lambda^{\circ}}^2 \int_0^{\delta}dq\,
{\mathcal K}_S^{\circ}(p,q) f^{\circ}(q).
\eeq
We find that the only nonzero eigenvalue is given by
\beq
\label{20}
\frac{1}{{\lambda^{\circ}}^2} \simeq \frac{\kappa^2}{2}
\log \frac{\delta}{\Delta},
\eeq
with the corresponding (normalized) eigenfunction
\beq
\label{eq:zevec}
f^{\circ}(p) = \frac{1}{\sqrt{\log (\delta/\Delta)}} \sqrt{\frac{p}{p^2 +
\Delta^2}} \theta(\delta - p).
\eeq
To obtain a perturbative solution to the full eigenvalue problem, we
introduce a complete set of orthonormal functions, $\langle p|m\rangle$
where $m=0,1,2,\ldots$ labels the eigenstate and $|0\rangle$ corresponds
to the state (\ref{eq:zevec}), {\it i.e.}~$\langle p|0\rangle=f^\circ(p)$.
Using the completeness relation $1=\sum_m|m\rangle\langle m|$ and the
fact that the states $|m\rangle$ for $m>0$ have zero eigenvalue under
the zeroth order kernel, the eigenvalue for the full kernel can then be
calculated to second order,
\beq
\label{15}
\frac{1}{\lambda^2} = \frac{1}{{\lambda^{\circ}}^2} +
\langle0|{\mathcal K}_S'|0\rangle
+ {\lambda^{\circ}}^2 \langle0|{\mathcal K}_S'^2|0\rangle
- {\lambda^{\circ}}^2 \langle0|{\mathcal K}_S'|0\rangle^2,
\eeq
where the inner product is defined by, for example,
\beq
\langle0|{\mathcal K}_S'|0\rangle = \int_0^\infty d q \int_0^\infty dp
f^{\circ}(p) {\mathcal K}_S'(p,q) f^{\circ}(q).
\eeq

To determine the sub-leading behaviour of (\ref{15}) relative to (\ref{20})
we are looking for contributions to the eigenvalue of ${\mathcal O}(1)$.
Since we are in
the weak coupling limit, expansion of the Bessel functions in (\ref{21})
quickly leads to the result,
\beq
\label{22}
\langle0|{\mathcal K}_S'|0\rangle \simeq {\mathcal O}\left(\frac{1}
{\log (\delta/\Delta)} \right),
\eeq
so that the first order correction is found to be insignificant.
This also implies that the last term of (\ref{15}) can be discarded.
We are then left with one term to calculate at second order.  Although the
zeroth order wavefunctions restrict the initial and final momenta to
lie in the range $[0,\delta]$, the intermediary momentum in
$\langle0|{\mathcal K}_S'^2|0\rangle$ (resulting from the convolution
of the kernel) is free to take all possible values $[0,\infty)$.
Although no small argument expansion of Bessel functions with the
intermediary momentum as an argument is allowed, the resulting integrals
can be performed exactly and give,
\beq
\label{23}
{\lambda^{\circ}}^2 \langle0|{\mathcal K}_S'^2|0\rangle = \frac{\kappa^2}{2}
\left( \log \left(\frac{2}{\delta a} \right) - \gamma + \frac{1}{4} \right).
\eeq

Substituting (\ref{20}), (\ref{22}) and (\ref{23}) into (\ref{15}) we
find that the dependence upon the arbitrary scale $\delta$ that
appears at zeroth order is cancelled by the second order correction,
as is required.  The leading and sub-leading behaviour of the eigenvalue
has thus been calculated to be
\beq
\frac{1}{\lambda^2} \simeq \frac{\kappa^2}{2}
\left(\log \left( \frac{2}{\Delta a} \right) - \gamma + \frac{1}{4} \right).
\eeq
{}From the condition $\lambda^2 =1$, the result (\ref{5}) is recovered.
%

\section{The Fredholm Equation for the Transition Temperature.}

We now return to the problem at hand and
consider an $SU(N)_c$ color gauge field coupling to $N_f$ flavors
of massless quarks with the Lagrangian density,
\begin{equation}
{\cal L}=-{1\over 4}F_{\mu\nu}^a{F_{\mu\nu}}^a-\bar\psi_f\gamma_\mu
(\partial_\mu-igA_\mu)\psi_f,
\label{eq:qcd}
\end{equation}
where $F_{\mu\nu}^a=\partial_\mu^{\vphantom{a}}A_\nu^a
-\partial_\nu^{\vphantom{a}}
A_\mu^a+gf^{abc}A_\mu^bA_\nu^c$ and $A_\mu=A_\mu^at^a$ with $t^a$ the
$SU(N)_c$ generator in its fundamental representation.  Since the
Lagrangian (\ref{eq:qcd}) is diagonal with respect to both flavor and
chirality, the corresponding indices will be dropped below.

\begin{figure}[t]
\epsfxsize 6cm
\centerline{\epsffile{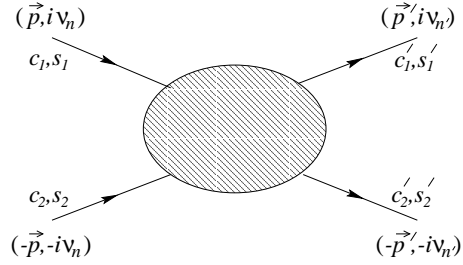}}
\bigskip
\caption{Proper vertex function,
$\Gamma_{s_1',s_2';s_1,s_2}^{c_1',c_2';c_1,c_2}(n',\vec p\,'|n,\vec p\,)$.}
\label{fig1}
\end{figure}

As in \cite{BLR99}, we shall study the proper vertex function, $\Gamma$,
corresponding to the scattering of two quarks with zero total energy
and momentum.  Our conventions are the same as in \cite{BLR99} but are
repeated here for clarity of presentation.  The Matsubara energies
$i\nu_n$ of individual ingoing and outgoing quarks are labelled by $n$
and $n'$, respectively.  Similarly, $\vec{p}$ and $\vec{p}\,'$ label
the ingoing and outgoing momenta.  Each of the superscripts $c_i$,
$i=1,2$, denote the color associated with each leg and the subscripts
$s$, which label the states above or below the Dirac sea, are either
$+$ or $-$.  Primed variables are outgoing and unprimed incoming.  The
ingoing quark pair are labelled by $s=(s_1,s_2)$ and $c=(c_1,c_2)$ and
the outgoing pair by $s'=(s_1',s_2')$ and $c'=(c'_1,c'_2)$.  This
labelling is illustrated in Fig.~\ref{fig1}.

\begin{figure}[t]
\epsfxsize 8cm
\centerline{\epsffile{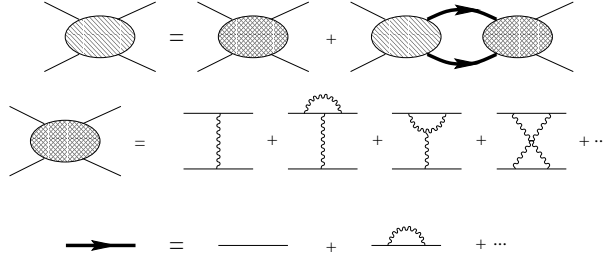}}
\bigskip
\caption{The Schwinger-Dyson equation.  $\Gamma$ is represented by single
hashed vertices and $\tilde\Gamma$ is represented by double hashed vertices.
The full quark propagator is represented by a solid line and the bare quark
propagator by a thin line.  Gluon propagators, including hard thermal loops,
are represented by curly lines.  The first two orders in the expansion of
$\tilde\Gamma$ and the full quark propagator are given.}
\label{fig2}
\end{figure}

The proper four-fermion vertex function satisfies a Schwinger-Dyson
equation (shown in Fig.~\ref{fig2}) which, with all
indices suppressed, may be written,
\begin{equation}
\label{eq:ie}
\Gamma(n^\prime,\vec p\,'\vert n,\vec p\,)
= \tilde\Gamma(n^\prime,\vec p\,' \vert n,\vec p\,)  +
{1\over\beta}\sum_m\int{d^3\vec q \over (2\pi)^3}
K(n^\prime,\vec p\,' \vert m,\vec q\,)
\Gamma(m, \vec q\, \vert n,\vec p\,),
\end{equation}
where $\tilde\Gamma$ represents the two quark irreducible vertex. The
kernel has the explicit form,
\beq
%
%
K_{s_1',s_2';s_1^{\vphantom{\prime}},s_2^{\vphantom{\prime}}}
^{c_1',c_2';c_1^{\vphantom{\prime}},c_2^{\vphantom{\prime}}}
(n^\prime,\vec p\,' \vert m,\vec q\,) = \tilde\Gamma
_{s_1',s_2';s_1^{\vphantom{\prime}},s_2^{\vphantom{\prime}}}
^{c_1',c_2';c_1^{\vphantom{\prime}},c_2^{\vphantom{\prime}}}
(n^\prime,\vec p\,' \vert m,\vec q\,)
S_{s_1}(m\vert\vec q\,)S_{s_2}(-m\vert-\vec q\,),
\eeq
where $S_s(n\vert\vec p\,)$ denotes the full quark propagator with
momentum $\vec p$ and Matsubara energy $i\nu_n={2\pi i\over\beta}
(n+{1\over 2})$.
In order to facilitate the partial wave analysis we found it convenient to
associate the Dirac spinors $u(\vec p\,)$ and $v(\vec p\,)$, which
satisfy the Dirac equations
$(\gamma_4p-i\vec\gamma\cdot\vec{p}\,)u(\vec p\,)=0$ and
$(\gamma_4p-i\vec\gamma\cdot\vec p\,)v(\vec p\,)=0$, to the quark-gluon
vertex instead of to the quark propagator.
Thus the vertices written in (\ref{eq:ie}) are of the form,
\begin{equation}
\Gamma_{s';s} = \overline U_{\gamma}(s_1',\vec p\,') \overline
U_\delta(s_2',-\vec p\,')\Gamma_{\gamma\delta,\alpha\beta}
U_{\alpha}(s_1,\vec p\,)U_{\beta}(s_2,-\vec p\,),
\end{equation}
with the vertex function $\Gamma_{\gamma\delta,\alpha\beta}$ given by
conventional Feynman rules and $U(+,\vec p\,)=u(\vec p\,)$ or
$U(-,\vec p\,)=v(-\vec p\,)$, respectively.  However, since the quarks
are massless, $\gamma_5 u = -u$ and $\gamma_5 v = -v$, to simplify
notation we may identify $U(s,\vec p\,)=u(s \vec p\,)$.

We complete this section of conventions and definitions with the
fermion and gluon propagators.  The zeroth order quark propagator
reads,
\beq
\label{quarkprop}
S_s(n\vert\vec p\,)=\frac{i}{i\nu_n-sp+\mu}.
\eeq
The radiative corrections to this propagator contribute to the pre-factor
${\mathcal C}_1$ in (\ref{TC}).  Such radiative corrections are depicted
in Fig.~\ref{fig2}.  This contribution has been calculated
in a previous publication, \cite{BLR99}, and the reader is directed there
for details.

We discuss the gauge invariance of this approach to the transition
temperature in section VI, but until then we shall work in the Coulomb
gauge where the gluon propagator takes the form,
\begin{eqnarray}
\label{eq:glue2}
D_{44}(\vec k , \omega) &=& D^E(\vec k , \omega),\\
\label{eq:glue1}
D_{ij}(\vec k , \omega) &=& D^M(\vec k, \omega)
(\delta_{ij}-{k_ik_j\over \vec k^2}), \\
\label{eq:glue3}
D_{4j}(\vec k , \omega) &=& D_{j4}(\vec k , \omega)=0.
\end{eqnarray}
In the presence of a Fermi sea, hard dense loops should be included
in the gluon propagator in the leading order.  While it is possible
that a magnetic mass of
order $T$ may exist, hard dense loops lead to Landau
damping which prevails at $\mu\gg k_BT$ \cite{Bellac}.  In general,
the screened electric and magnetic propagators take the form,
\begin{eqnarray}
\label{eq:glue4}
D^E(\vec k , \omega) &=& \frac{-i}{ \vec{k}^2 + \sigma^E(\vec{k},
\omega)},\\
\label{eq:glue5}
D^M(\vec k, \omega) &=& \frac{-i}{\vec{k}^2 + \omega^2 +
\sigma^M(\vec{k},\omega)},
\end{eqnarray}
In the region $k \ll \mu$ and $\omega \ll \mu$, both
$\sigma^E(\vec{k},\omega)$ and $\sigma^M(\vec{k},\omega)$ depend only
upon the ratio $x = \omega/|\vec{k}|$, $x \in (-\infty,\infty)$.
Scaling out the Debye mass,
\begin{equation}
m_D^2={N_f g^2\over \pi^2} \int_0^\infty dq\,q{1\over e^{\beta(q-\mu)}+1}
\simeq \frac{N_f g^2 \mu^2}{2 \pi^2},
\end{equation}
by writing $\sigma^E(\vec{k},\omega)
=m_D^2 f_E^{\vphantom{2}}(x)$ and $\sigma^M(\vec{k}, \omega)
=m_D^2 f_M^{\vphantom{2}}(x)$, we find the dimensionless screening functions,
\begin{eqnarray}
\label{eq:F}
f_E(x) &=&   \left[ 1 - x \tan^{-1} \left( \frac{1}{x}
\right) \right],\\
\label{eq:G}
f_M(x) &=& \frac{x}{2} \left[ (1+x^2) \tan^{-1} \left(
\frac{1}{x} \right) - x \right].
\end{eqnarray}
Analysis of the functions $f_E(x)$ and
$f_M(x)$ shows that the dangerous region for infrared
divergences to originate in the gluon propagator is at small $x$,
corresponding to energy and momentum transfer $\omega \ll k \ll \mu$.
In this region the gluon self-energy functions are well approximated by
\beqa
\label{escreen}
f_E(x) &\simeq& 1, \\
\label{mscreen}
f_M(x) &\simeq& \frac{\pi}{4} |x|,
\eeqa
indicating that, while the Coulomb interaction is strongly screened by
the Debye length, the magnetic interaction is poorly screened.  It is
for this reason that the attractive interaction responsible for the
pairing instability is dominated by single collinear magnetic gluon
exchange.  Furthermore, $f_E(x)$ is a decreasing function of $x$ and
$f_M(x)$ is an increasing function of $x$, which obey the following
inequalities for all $x$:
\beqa
f_E(x) &\leq& 1,\\
f_M(x) &\leq& \frac{\pi}{4}|x|.
\eeqa
For $x \gg 1$ we have,
\beqa
f_E(x) &\simeq& \frac{1}{3x^2},\\
f_M(x) &\simeq& \frac{1}{3}.
\eeqa

Since $\Gamma$ corresponds to di-quark scattering it can be decomposed
into irreducible representations of $SU(N)_c$ by either symmetrization
or antisymmetrization among the initial and final color indices.  We
decompose $\Gamma$ into its symmetric and anti-symmetric components,
\begin{eqnarray}
\Gamma_{s^\prime,s}^{c^\prime, c}(n^\prime,\vec p\,'|n,\vec p\,)
&=& \sqrt{2}\;\delta^{c_1^{\vphantom{\prime}}(c_1'}
\delta^{c_2')c_2^{\vphantom{\prime}}}
\Gamma_{s^\prime,s}^S(n^\prime,\vec p\,', |n,\vec p\,) \\ \nonumber
&+& \sqrt{2}\;\delta^{c_1^{\vphantom{\prime}}[c_1'}
\delta^{c_2']c_2^{\vphantom{\prime}}}
\Gamma_{s^\prime,s}^A(n^\prime,\vec p\,',|n, \vec p\,),
\label{eq:symanti}
\end{eqnarray}
where $[\cdots]$ and $(\cdots)$ denote antisymmetrization and
symmetrization with weight one respectively (and likewise for
$\tilde\Gamma$). Since the Fermi surface has a pairing
instability in the presence of even an arbitrarily weak attractive
interaction we need only focus on the attractive antisymmetric channel.

Proceeding with the partial wave analysis, we expand
$\Gamma_{s^\prime,s}^A(n^\prime,\vec p\,'|n,\vec p\,)$ in
terms of Legendre polynomials,
\begin{equation}
\Gamma_{s^\prime,s}^A(n^\prime,\vec p\,'|n,\vec p\,)
=\sum_J\gamma_{s^\prime,s}^J(n^\prime,p^\prime \vert n,p)
P_J(\cos\theta).
\label{eq:legend}
\end{equation}
Using a similar expression for
$\tilde\Gamma_{s^\prime,s}^A(n^\prime,\vec p\,' \vert n,\vec p\,)$, we find
from (\ref{eq:ie}) the Fredholm equation satisfied by
$\gamma_{s^\prime,s}^J(n^\prime, p^\prime | n ,p)$:
\beq
\label{7}
 \gamma_{s^\prime,s}^J(n^\prime, p^\prime|n,p) =
\tilde\gamma_{s^\prime,s}^J (n^\prime,p^\prime|n,p)
+{1\over\beta}\sum_{m,s^{\prime\prime}}
\int_0^\infty dq\,K_{s^\prime,s^{\prime\prime}}^J(n^\prime,p^\prime |m,q)
\gamma_{s^{\prime\prime},s}^J(m,q \vert n,p),
\eeq
where the kernel $K_{s',s}^J$ has the form,
\begin{equation}
K_{s^\prime,s}^J(n^\prime, p^\prime |n,p)
={p^2\tilde\gamma_{s^\prime,s}^J(n^\prime, p^\prime |n,p)
\over 2\pi^2(2J+1)}
S_{s_1}(n\vert p)
S_{s_2}(-n\vert p).
\label{kernel}
\end{equation}

This integral equation can be cast into a general eigenvalue problem
in the way outlined for the toy model in section II.  To do so, we first
note that the formal solution of (\ref{7}) involves the Fredholm determinant
\beq
{\mathcal D} = \det (1-K) = \prod_j (1 - \lambda_j^{-2}),
\eeq
where the eigenvalues, $\{\lambda_j^2\}$, are defined by the solutions of,
\beq
\label{eq:gen}
f_{s}(n,p) = \frac{\lambda^2}{\beta} \sum_{n',s'} \int_0^{\infty} dp'
K_{s',s}^J (n',p'|n,p) f_{s'}(n',p').
\eeq
Since ${\cal D}$ appears in the denominator in the inversion of the
Fredholm equation, it in fact governs the pole in the solution that
represents the formation of a bound state of quarks at the transition
point to the superphase. For weak coupling, we may identify the pairing
temperature and the phase coherence temperature.

At sufficiently high temperature all of the set of eigenvalues
$\{\lambda_j^2\}$ are greater than one, so that ${\mathcal D} \neq 0$
and there is no instability---the theory is in the normal phase.
As the temperature is reduced, we find the transition temperature to
the superphase is that at which the smallest of $\{\lambda_j^2\}$
reaches one.  If we label the solutions so that the smallest eigenvalue
is $\lambda_0^2$, then the transition temperature is obtained from the
inversion of the condition, $\lambda_0^2(T_C,g,\mu) = 1$.

In order to proceed with a second order perturbative calculation, it is
easiest to work with a symmetric kernel.  In the general eigenvalue
problem defined by (\ref{eq:gen}) we have a kernel $K_{s',s} \equiv
K_{s'_1s'_2,s_1^{\vphantom{\prime}}s_2^{\vphantom{\prime}}}$, where
the leading order behaviour
only occurs for particle states (that can see the Fermi sea), namely for
$s = (+,+)$ and $s' = (+,+)$.  In general, however, the
solution $f_{++}$ is necessarily coupled to solutions $f_{+-}$, $f_{-+}$ and
$f_{--}$ by the kernels $K_{+-,++}$, $K_{-+,++}$ and $K_{--,++}$,
respectively.  On the other hand, since `$-$' represents a quark below the
Dirac sea, its corresponding propagator has no pole for momentum at the
Fermi sea, $p=\mu$, as may be seen from (\ref{quarkprop}).  In fact, the
absence of this pole is sufficient to remove two orders of the
logarithm and hence the cross-channels make no contribution to either the
leading or the subleading behaviour.  In the following we shall focus
on the kernel $K=K_{++,++}$ and drop all particle/antiparticle notation.

The kernel $K$ is symmetric under simultaneous switch of the sign of the
Matsubara labels $n^\prime$ and $n$, {\it i.e.}, $K(n',p'\vert n, p)=
K(-n',p' \vert -n, p)$.  As a result, the eigenfunctions $f(n,p)$ must
then be either even or odd with respect to $n$. Since only the even
functions pick up the singularities on the Fermi surface we shall
concentrate on them and replace the sum $\sum_{n=-\infty}^\infty$ by
$\sum_{n=0}^\infty$.

Symmetrization of the problem proceeds by identifying
\beq
f(n,p) = \frac{\sqrt{\nu_n^2 + (p - \mu)^2}}{p} h(\nu_n,p).
\eeq
Then from (\ref{eq:gen}) and (\ref{kernel}) and using
$h(\nu_n,p) = h(-\nu_n,p)$ we find,
\beq
\label{eq:hint}
h(\nu_n,p) = \frac{\lambda^2}{\beta} \sum_{n'=0}^{\infty}
\int_0^{\infty} dp' K_S^J(n',p'|n,p) h(\nu_{n'},p'),
\eeq
where the symmetrized kernel, $K_S^J$, is written,
\beq
\label{326}
K_S^J(n' , p'|n,p) = - \frac{pp'}{2 \pi^2 (2J + 1)}
\left\{\frac{\tilde{\gamma}^J(n' , p'|n,p) + \tilde{\gamma}^J(-n' , p'|n,p)}
{\sqrt{(\nu_n^2 + (p - \mu)^2)(\nu_{n'}^2 + (p' - \mu)^2)}}\right\}.
\eeq
This will be the starting point for the perturbative calculation of the
next section.

\section{Perturbative Calculation of the Transition Temperature.}
In this section we shall concentrate upon the calculation of the
transition temperature for a longitudinal pair in the $s$-wave
channel, $J=0$.  The discussion of the transition temperatures
corresponding to higher angular momenta longitudinal and transverse
pairings will be deferred until section V.  To zeroth order in the
perturbative expansion, we start with the leading order solution
obtained previously in
\cite{son1998,SWc,pisarski1999b,BLR99,hs,pisarski1999c}.  In the
following subsections we shall define and evaluate a
perturbation series to uncover the subleading behaviour.

For one gluon exchange, incorporating hard dense loops in
the bare quark propagators, we find (in the $K_{++,++}$ channel),
\beqa
\label{kfull}
\tilde\gamma^{J=0}(n^\prime,p^\prime |n,p)&=&
-\frac{g^2}{8}\left( 1 + \frac{1}{N} \right)\int_{-1}^{1} d(\cos \theta) \\
& & \qquad \times \left[
\frac{3-\cos\theta - (1+\cos\theta) \frac{(p-p')^2}{|\vec{p}-\vec{p}\,'|^2}}
{(\nu_n - \nu_{n'})^2 + |\vec{p} - \vec{p}\,'|^2 + m_D^2f_M(x)}
+ \frac{1 + \cos \theta}
{|\vec{p} - \vec{p}\,'|^2 + m_D^2f_E(x)}\right], \nonumber
\eeqa
where $x=(\nu_n-\nu_{n'})/|\vec p-\vec p\,'|$.  The first term in the
integrand arises from magnetic gluon exchange and the second from
electric gluon exchange.  Further discussion of this vertex is contained
in appendix A.  To obtain the leading double logarithmic
behaviour we see that the infrared-sensitive region is in the forward
direction, $\theta\approx0$, so that $\cos\theta$ may be approximated
by one in the numerator of the integrand.  Furthermore, we discard the
energy transfer term, $(\nu_n-\nu_{n^\prime})^2$, in the denominator
of the magnetic propagator and approximate the self-energy functions,
$f_E(x)$ and $f_M(x)$, by their asymptotic forms
(\ref{escreen}) and (\ref{mscreen}) for small $x$. These
approximations, prevalent in the literature, over-estimate both the
electric and magnetic screening and thus under-estimate the attraction
between quarks.  How this correlates with the transition temperature
is hard a priori to quantify.  Although a weaker attraction should
lead to a lower transition temperature and smaller condensates we
shall see that, in fact, none of the corrections to these leading order
approximations discard any next-to-leading behaviour of the
eigenvalue.

With these approximations, the integration over $\cos\theta$ becomes trivial,
\beq
\label{approx}
\tilde\gamma^{J=0}(n,p|n^\prime,p^\prime)\simeq -\frac{g^2}{12pp^\prime}
\left( 1 + \frac{1}{N} \right) \left[
\log \frac{8\mu^3}{|p-p^\prime|^3+\frac{\pi}{4}m_D^2|\nu_n-\nu_{n^\prime}|}
+{3\over2} \log\frac{4\mu^2}{m_D^2} \right].
\eeq
To leading order, we may also neglect the momentum exchange under the
first logarithm.  Although this produces a logarithmic singularity for
$n=n'$ which is potentially dangerous, we shall see that in fact it is
well behaved.  Combining the electric and magnetic parts we find,
\beq
\label{ap0}
\tilde\gamma^{J= 0}(n',p'|n,p)\simeq -\frac{g^2}{12pp'}
\left( 1 + \frac{1}{N} \right)
\log\frac{1}{|\hat\nu_n-\hat\nu_{n^\prime}|},
\eeq
where
\beq
\label{eq:nuscale}
\hat\nu={g^5\over256\pi^4}\left({N_f\over2}\right)^{5\over 2} {\nu\over\mu}.
\eeq
%
%
Using the Ansatz,
\beq
h^\circ(\nu_n,p)=\frac{\chi(\nu_n)}{\sqrt{\nu_n^2+(p-\mu)^2}},
\eeq
and substituting into (\ref{eq:hint}), we find that the leading order
eigenvalue problem becomes,
\beq
\label{eq:loevp}
\chi_j(\nu_n) = {\lambda_j^{\circ}}^2\frac{g^2}{24\pi^2}
\left(1+\frac{1}{N}\right) \frac{1}{\beta} \sum_{n'=0}^{\infty}
\int_0^\infty dp'  K^\circ (n',p'|n,p) \sqrt{\frac{\nu_n^2 +
(p - \mu)^2}{\nu_{n'}^2 + (p' - \mu)^2}}
\chi_j(\nu_n')
\eeq
where the kernel focuses upon the most divergent region of the Matsubara
energies, and we impose a cut, $\hat\nu < \hat\delta$, analogous to that
used for the toy model in section II:
\beq
\label{dl}
K^\circ(n',p'|n,p)=\left\{\frac{
\log\frac{1}{|\hat{\nu}_{n^{\vphantom{\prime}}} - \hat{\nu}_{n'}|}
+\log \frac{1}{|\hat{\nu}_{n^{\vphantom{\prime}}} + \hat{\nu}_{n'}|}}
{\sqrt{(\nu_n^2 + (p - \mu)^2) (\nu_{n'}^2 + (p' - \mu)^2)}} \right\}
\theta(\hat\delta - \hat\nu_n) \theta (\hat\delta - \hat\nu_{n'}).
\eeq
for $n\neq n^\prime$ and
\beq
\label{eq:neqnp}
K^\circ(n,p'|n,p)=3\left\{\frac{
\log\frac{p+p^\prime}{|p-p^\prime|}}
{\sqrt{(\nu_n^2 + (p - \mu)^2) (\nu_n^2 + (p' - \mu)^2)}} \right\}
\theta(\hat\delta - \hat\nu_n).
\eeq
To calculate the leading order behaviour we move to the continuum
limit,
\begin{equation}
\label{eq:matsum}
{1\over\beta}\sum_{n'=0}^\infty \to
{1\over2\pi}\int_{\epsilon\over2}^\infty d\nu',
\end{equation}
where $\epsilon=2\pi k_BT$, and perform the momentum integral for $p'$,
\begin{equation}
\label{eq:pint}
\int_{p'\approx\mu} dp'{1\over[\nu_{n'}^2+(p'-\mu)^2]} \to {\pi\over\nu_{n'}}
\end{equation}
(assuming $\nu_{n'} \ll \mu$).  Furthermore, the kernel $K^\circ$,
containing both $-\log|\hat\nu-\hat\nu'|$ and $-\log|\hat\nu+\hat\nu'|$,
can be treated using the same approximation as employed by Son \cite{son1998}.
This amounts to replacing the sum of these terms,
\beq
\label{sonsapprox}
\log \frac{1}{|\hat{\nu} - \hat{\nu}'|} + \log \frac{1}{|\hat{\nu} +
\hat{\nu}'|} \approx 2 \log \frac{1}{\hat\nu_>},
\eeq
where $\hat\nu_>= \max(\hat\nu,\hat\nu^\prime)$.  Although this
approximation may appear
drastic, in fact it captures the behaviour remarkably well.  We will
show in the next section that the correction to the full kernel due to
this approximation does not contribute a linear logarithm but that the
continuous-discrete energy approximation does.

In the continuum limit we incorporate (\ref{eq:matsum}), (\ref{eq:pint})
and (\ref{sonsapprox}) into (\ref{eq:loevp}), and are left with the
integral equation,
\beq
\label{sons}
\chi_j(\nu) = {\lambda_j^\circ}^2 \frac{g^2}{24\pi^2}
\left(1 + \frac{1}{N} \right)
\int_{\frac{\hat\epsilon}{2}}^{\hat\delta} \frac{d\hat{\nu'}}{
\hat{\nu}'} \log \frac{1}{\hat{\nu}_>} \chi_j(\nu'),
\eeq
where the ultraviolet energy cut-off $\hat\delta \sim 1$ is analogous to the
cut made in momentum for the toy model in (\ref{cut}).  Note that the
rescaling of $\nu$ in (\ref{eq:nuscale}) results in a re-scaled infrared
cutoff, $\hat\epsilon = (N_f/2)^{5/2}\frac{g^5 k_B T}{128 \pi^3 \mu}$.
While (\ref{sons}) may be solved by Son's method for arbitrary cutoff
$\hat\delta$, the results take the simplest form when $\hat\delta=1$.  In
this case, we find the solutions
\begin{equation}
\label{sol}
{\chi_j}(\nu) = \theta(\hat\delta-\hat\nu)
\frac{2}{\sqrt{\log 2/\hat\epsilon}}
\sin\left[\left(j + \frac{1}{2}\right) \pi
\frac{\log 1/\hat\nu}{\log 2/\hat\epsilon}\right],
\end{equation}
with eigenvalues
\begin{equation}
\label{eq:evals}
\frac{1}{\lambda_j^\circ} = \frac{g}{\left( 2j + 1 \right) \pi^2}
\sqrt{\frac{N+1}{6N}}
\log \frac{2}{\hat\epsilon}\qquad\hbox{(for $\hat\delta=1$)}.
\end{equation}
The functions $h_j(\nu,p)$ form a orthonormal set with respect to
the continuum approximation, {\it i.e.}
\beq
\int_{\frac{\epsilon}{2}}^\delta\frac{d\nu}{2\pi}
\int_0^\infty dp\, h_i(\nu,p)h_j(\nu,p)=\delta_{ij},
\eeq
where we have dropped the contribution from the lower limit of $p$
integration.  This approximation will be adapted in the subsequent work.

For $\hat\delta\ne1$, the eigenvalues are instead obtained by solution of
the transcendental equation (\ref{eq:stran}), resulting in
\begin{equation}
\label{eq:cevals}
\frac{1}{\lambda_j^\circ} = \frac{g}{\left( 2j + 1 \right) \pi^2}
\sqrt{\frac{N+1}{6N}}
\left( \log \frac{2}{\hat\epsilon} + \alpha +\cdots\right),
\end{equation}
where
\beq
\alpha = -\frac{1}{3}[(j + {\textstyle\frac{1}{2}})\pi]^2
\frac{\log^3 1/\hat\delta}{\log^2 2/\hat\epsilon}.
\eeq
As may be seen from (\ref{eq:matsum}), the fermion Matsubara sum in
the finite temperature formalism automatically provides a natural
infrared cutoff to the continuum eigenvalue problem.  Also, the
dependence on the ultraviolet cutoff, $\hat\delta$, is beyond the order
to which we are interested in, unlike in the toy model. This allows
$\alpha$ to be ignored, so that (\ref{sol}) and (\ref{eq:evals}) are
sufficient in the subsequent calculations.  More discussion on the
dependence of the results on the ultraviolet cutoff is given in section VI.

One concern when defining the general eigenvalue problem is that,
with a non-hermitian kernel (\ref{kernel}), in general the smallest
eigenvalue defined by (\ref{eq:gen}) may be complex.  Since the
leading behavior in coupling of the transition temperature is solely
determined by one gluon exchange, in analyzing this question we shall
focus our attention on the corresponding eigenvalue problem with a
kernel defined by (\ref{kernel}) and (\ref{kfull}).  In this case, by
direct calculation we have shown that the leading order of the
eigenvalue is in fact real.  We shall briefly show, further, that as
long as it is not degenerate then the smallest eigenvalue must be
real to all orders.  To do so, we examine the characteristic polynomial
of the eigenvalue problem, ${\cal P}(\rho) = \det P(\rho)$ where
\beq
%
%
P_{s'_1s'_2,s_1^{\vphantom{\prime}}s_2^{\vphantom{\prime}}}(\rho)
\equiv K_{s'_1s'_2,s_1^{\vphantom{\prime}}s_2^{\vphantom{\prime}}}
- \rho\,\delta_{s'_1s_1^{\vphantom{\prime}}}
\delta_{s'_2s_2^{\vphantom{\prime}}}.
\eeq
To investigate the coefficients of ${\cal P}(\rho)$ we first consider
$\rho$ to be real.  In this case, we find the relation
\beq
P_{s'_1s'_2,s_1^{\vphantom{\prime}}s_2^{\vphantom{\prime}}}(\rho)
= (-)^{\frac{s_1^{\vphantom{\prime}} - s_2^{\vphantom{\prime}}}{2}
+ \frac{s'_1 - s'_2}{2}}
P_{s'_2s'_1,s_2^{\vphantom{\prime}}s_1^{\vphantom{\prime}}}^*(\rho),
\eeq
by explicit examination of (\ref{kernel}) for the one gluon exchange
in different channels, tabulated in appendix A together with the property
that
\beq
S_{s_1}(n|p)S_{s_2}(-n|p)=S_{s_2}^*(n|p)S_{s_1}^*(-n|p)
\eeq
for a bare quark propagator.
Therefore it follows that ${\rm det}P(\rho) = {\rm det}P^*(\rho)$, so
that the coefficients of the characteristic polynomial are all real.
Continuing $\rho$ to be complex, the real coefficients of the
characteristic polynomial imply that the eigenvalues determined by the
characteristic equation, ${\cal P}(\rho) = 0$, are either real or
appear in complex conjugate pairs.  Therefore, if the smallest
eigenvalue is not degenerate it is real.  Although this argument is
specifically for the problem defined by one gluon exchange, we expect
similar reasoning to extend to the general eigenvalue problem.

\subsection{The Perturbation Series.}

Given the above leading order solution, we now turn to the issue of
perturbative corrections. To simplify the notation, we remove
non-essential factors by rescaling the kernel and eigenvalues by
\beq
K_S^J(n^\prime,p^\prime|n,p)=\frac{g^2}{24\pi^2}\left(1+\frac{1}{N}\right)
\widetilde K_S^J(n^\prime,p^\prime|n,p),
\eeq
and
\beq
\frac{1}{\lambda_j^2}=\frac{g^2}{24\pi^2}\left(1+\frac{1}{N}\right)
\frac{1}{\widetilde\lambda_j^2}.
\eeq
Then to the leading order, we have
$\widetilde\lambda_0=\widetilde\lambda_0^\circ$ with
\beq
\frac{1}{\widetilde\lambda_j^\circ} = \frac{1}{(j+\frac{1}{2})\pi}
\log \frac{2}{\hat\epsilon}.
\eeq

One concern, that eigenvalues of the
general problem may be complex, was discussed in the previous
sub-section where it was found that the smallest eigenvalue---the
eigenvalue of interest in defining the transition temperature---is
real if it is not degenerate.  Another concern in defining a
perturbation series with discrete energy is that the leading order
solutions $h_j^{\circ}(\nu_n,p)$ [obtained from the substitution $\nu
\rightarrow \nu_n$ in (\ref{sol})] are not orthogonal and cannot be
directly used as a basis for the perturbative expansion.  However,
each of the solutions may be rotated slightly so that the resulting set forms
a complete orthonormal basis.  In a more convenient notation, in which
$\langle \nu_n, p|j_c\rangle = h_j^{\circ}(\nu_n,p)$, while
$\langle \nu_n, p|j \rangle = {\langle j_c|j_c\rangle}^{-1/2}
h_j^{\circ}(\nu_n,p)$  we can use Gram-Schmidt
decomposition to write down an orthonormal basis,
\beq
\label{orthobasis}
|\tilde{j} \rangle = \frac{|j\rangle - \sum_{i=0}^{j-1} | \tilde{i} \rangle
\langle \tilde{i} | j \rangle}
{\sqrt{1 - \sum_{i=0}^{j-1} \langle j | \tilde{i}
\rangle \langle \tilde{i} | j \rangle}},
\eeq
where all $|j\rangle$ are normalized with respect to the discrete sum
over Matsubara energy,
$|\tilde{0}\rangle = |0\rangle$ and other $|\tilde{j}\rangle$ for
$j > 0$ are found inductively. The norm $\langle j_c|j_c\rangle$ is
examined in appendix B along with the inner product
$\langle j_c|j_c^\prime\rangle$ and some matrix elements of
the perturbative kernel. In this basis, we define the
zeroth order operator with discrete energy,
\beq
\label{zeroop}
{\mathcal K}_0 = \sum_i \widetilde\lambda_i^{\circ\,^{-2}} |\tilde{i}\rangle
\langle \tilde{i} |.
\eeq
This operator has exactly the eigenvalues (\ref{eq:evals}) by
definition: ${\mathcal K}_0$ is diagonalized by $|\tilde{j}\rangle$
and so $|\tilde{j}\rangle = \widetilde{\lambda}_j^{\circ\,2} {\mathcal K}_0
|\tilde{j}\rangle$.  Thus ${\cal K}_0$ captures the desired leading
order behavior of the full kernel at hand.  In appendix C it is shown
that in the continuum limit this operator in fact reduces exactly to the
leading order kernel
(\ref{dl}) that contains the double logarithmic behaviour.

We now write the full kernel of (\ref{eq:hint}) in terms of this zeroth order
operator and two correction terms,
\beq
\widetilde K_S^{J=0} = {\mathcal K}_0 + {\mathcal K}_1 + {\mathcal K}_2,
\eeq
where
\begin{equation}
{\mathcal K}_1 = K^\circ - {\mathcal K}_0
\end{equation}
contains both the effects of discrete to continuous energy corrections as
well as corrections due to the approximation (\ref{sonsapprox}) and
\begin{equation}
{\mathcal K}_2 = \widetilde K_S^{J=0} - K^\circ
\end{equation}
accounts for corrections between $K^\circ$ and the full kernel
$\widetilde K_S^{J=0}$.
In particular, the sub-leading behaviour discarded by working in the
continuum limit to derive (\ref{sol}) and (\ref{eq:evals}) is contained
in ${\cal K}_1$.

Following standard perturbation theory, the eigenvalue problem can be
calculated up to second order, which proves sufficient to obtain all the
next-to-leading order behaviour required.  To define our perturbative
expansion, we note that, to form a complete set, there are actually two
types of wavefunction at zeroth order.  The first type are determined by
the kernel $K^\circ$ (and hence appear in the definition of ${\cal K}_0$),
are denoted by $|\tilde{l}\rangle$ (which includes
$|\tilde{0}\rangle$) and have eigenvalues $\widetilde{\lambda}_l^{\circ\,-2}$.
These wavefunctions are constrained to have support only in the domain of
$K^\circ$ ({\it i.e.}~$\hat\nu\in [\hat\epsilon/2,\hat\delta]$), and are zero
outside.  The second type of wavefunction are not yet determined; to
distinguish them from the first type they are denoted by $|m)$ and have
eigenvalues $\widetilde\lambda_m^{-2}=0$.  Together, these solutions give us
the completeness relation,
\beq
1 = | 0 \rangle \langle 0 | + \sum_{j \neq 0} | \tilde{j} \rangle \langle
\tilde{j}| + \sum_m | m )(  m |.
\eeq
Using second order perturbation theory and the completeness relation, we
find
\begin{eqnarray}
\label{pertseries}
{1\over\widetilde\lambda_0^2} &=& {1\over{\widetilde{\lambda}_0^{\circ\,^2}}}
+ \langle\tilde0|({\cal K}_1+{\cal K}_2)|\tilde0\rangle\nonumber\\
&&\qquad
+\sum_{l\ne0} {\langle\tilde0|({\cal K}_1 + {\cal K}_2)|\tilde{l}\rangle
\langle \tilde{l} |({\cal K}_1 + {\cal K}_2)|\tilde0\rangle
\over{1\over{\widetilde\lambda_0^{\circ\,^2}}}
-{1\over{\widetilde\lambda_l^{\circ\,^2}}}}
+\sum_{m} {\langle\tilde0|({\cal K}_1 + {\cal K}_2)|m)
(m|({\cal K}_1 + {\cal K}_2)|\tilde0\rangle
\over{{1\over{\widetilde\lambda_0^{\circ\,^2}}}-0}}\nonumber \\
&=& {1\over{\widetilde{\lambda}_0^{\circ\,^2}}}
+ \langle \tilde{0} |({\cal K}_1 + {\cal K}_2)| \tilde{0} \rangle
+ {\widetilde\lambda_0^{\circ\,^2}}\langle\tilde0|
({\cal K}_1+{\cal K}_2)^2|\tilde0\rangle
- {\widetilde\lambda_0^{\circ\,^2}}\langle\tilde0|
({\cal K}_1+{\cal K}_2)|\tilde0\rangle^2\\
\nonumber
&&\qquad + \sum_{l \neq 0}
\left(\frac{1}{\frac{1}{{\widetilde\lambda_0^{\circ\,^2}}}
- \frac{1}{{\widetilde\lambda_l^{\circ\,^2}}}} -
  {\widetilde\lambda_0^{\circ\,^2}} \right)
\langle \tilde{0}|({\cal K}_1 + {\cal K}_2)|\tilde{l} \rangle
\langle \tilde{l} |({\cal K}_1 + {\cal K}_2)| \tilde{0} \rangle,
\end{eqnarray}
which may be compared to the analogous expression for the toy model,
(\ref{15}).

We now proceed to evaluate (\ref{pertseries}) and to resolve the
eigenvalue problem to next-to-leading order in the logarithmic
behaviour.  This will not only determine the exponential and the
pre-exponential factor of the transition temperature to the superphase
of QCD to leading order in the weak coupling regime but also allow us
to identify the physical origins of the contributions.  Since the
leading order is double logarithmic, the sub-leading constant
pertaining to the logarithm is solely determined by linearly
logarithmic terms.  Contributions of ${\mathcal O}(1)$ may be dropped.

\subsection{First order perturbation.}
The first order term in (\ref{pertseries}) is evaluated in two parts, the
first of which is the expectation of ${\mathcal K}_1$.  This itself
can be separated into two parts: ${\mathcal K}_1^{(a)}$ containing
corrections due to the approximation of the log terms given in
(\ref{sonsapprox}), and
${\mathcal K}_1^{(b)}$ containing significant continuous-discrete
energy corrections,
\beqa
{\mathcal K}_1 &=& {\mathcal K}_1^{(a)} + {\mathcal K}_1^{(b)}, \\
\label{eq:k1a}
{\mathcal K}_1^{(a)} &=& \theta(\hat\delta-\hat\nu_n)
\theta(\hat\delta-\hat\nu_{n'})\Big[
\frac{ \log \frac{1}{|\hat{\nu}_{n^{\vphantom{\prime}}} - \hat{\nu}_{n'}|}
+ \log \frac{1}{|\hat{\nu}_{n^{\vphantom{\prime}}} + \hat{\nu}_{n'}|}
-2 \log \frac{1}{\hat{\nu}_>}}
{\sqrt{(\nu_n^2+(p-\mu)^2)(\nu_{n^\prime}^2+(p^\prime-\mu)^2)}}
(1-\delta_{nn^\prime})\nonumber \\
&&\kern3.2cm+\frac{ 3\log \frac{p+p^\prime}{|p-p^\prime|}
-2 \log \frac{1}{\hat{\nu}_>}}
{\sqrt{(\nu_n^2+(p-\mu)^2)(\nu_{n^\prime}^2+(p^\prime-\mu)^2)}}
\,\delta_{nn^\prime}\Big], \\
\label{eq:kern1b}
{\mathcal K}_1^{(b)} &=& \theta(\hat\delta-\hat\nu_n)
\theta(\hat\delta-\hat\nu_{n'})
\frac{2 \log \frac{1}{\hat{\nu}_>}}
{\sqrt{(\nu_n^2+(p-\mu)^2)(\nu_{n^\prime}^2+(p^\prime-\mu)^2)}}
-{\cal K}_0.
\eeqa
In appendix D it is shown that the contribution from ${\mathcal K}_1^{(a)}$
may be ignored,
\beq
\langle \tilde{0}|{\mathcal K}_1^{(a)}|\tilde{0}\rangle =  {\mathcal O}(1).
\eeq
Significant continuous-discrete energy corrections do, however, arise
from the expectation of ${\mathcal K}_1^{(b)}$.  This expression is
evaluated using Zeta function techniques, and details of the calculation
are given in appendix B.  In total, we find the contribution from the first
term to be,
\beq
\label{cont-disc}
\langle \tilde{0}| {\mathcal K}_1 | \tilde{0} \rangle =
{4\over\pi^2}\left[2
(\gamma + \log 2 ) \log{2\over\hat\epsilon} + {\mathcal O}(1)\right].
\eeq

The second term to be evaluated at first order, ${\mathcal K}_2$,
contains the correction between the full kernel for one gluon exchange
and the approximate kernel containing $K^\circ$.  The first order
perturbation of ${\cal K}_2$ is given by,
\beqa
\label{k2}
\langle\tilde 0|{\cal K}_2|\tilde 0\rangle &= &\frac{1}{\beta^2}
\sum_{n,n^\prime} \int_0^\infty dp\int_0^\infty dp^\prime
\frac{\widetilde K_S^{J=0}(n',p'|n,p)-K^\circ(n',p'|n,p)}
{\sqrt{(\nu_n^2+(p-\mu)^2)(\nu_{n^\prime}^2+(p^\prime-\mu)^2)}}\chi_0(\nu_n)
\chi_0(\nu_{n^\prime})\\ \nonumber
&=& \int_{\frac{\epsilon}{2}}^{\delta} \frac{d\nu}{2\pi}
\int_{\frac{\epsilon}{2}}^{\delta} \frac{d\nu^\prime}{2\pi}
\chi_0(\nu)\chi_0(\nu^\prime)\int_0^\infty dp\int_0^\infty dp^\prime
\frac{\widetilde K_S^{J=0}-K^\circ}{\sqrt{(\nu^2+(p-\mu)^2)(\nu^{\prime2}+
(p^\prime-\mu)^2)}}.
\eeqa
To produce a linear logarithmic term, or in other words to knock out
one logarithmic power relative to the leading order, both $\nu$ and
$\nu^\prime$ should be $\sim k_BT$. If either $\nu$ or $\nu^\prime$
becomes large in comparison with $k_BT$ it will bring the integrand
away from the corresponding Fermi sea and therefore away from the
sensitive region of the gluon propagators so two powers of a logarithm
will be eliminated. For this reason, the integrations over $p$ and
$p^\prime$ are dominated at the poles $p-\mu=\pm i \nu$ and
$p^\prime-\mu=\pm i \nu^\prime$ and the energy transfer square in the
denominator of the magnetic gluon propagator can be disregarded.

If the behaviour of (\ref {k2}) is in fact linearly logarithmic it would be
tied to the constant pertaining to the logarithms of $K^\circ$.  Thus
(\ref{k2}) demands a closer look and we start with the partial wave integrals
of single electric and magnetic gluon interaction that combine to give
$\tilde{\gamma}^{J=0}$,
\beq
\tilde{\gamma}^{J=0} = -\frac{g^2}{2} \left( 1 + \frac{1}{N} \right) \left[
\tilde\gamma^{(M)}+\tilde\gamma^{(E)} \right].
\eeq
In the case $p=p'=\mu$,
\beq
\label{k2m}
\tilde\gamma^{(M)} \simeq \frac{1}{4}\int_{-1}^1d\cos\theta
\frac{3-\cos\theta}{2\mu^2(1-\cos\theta)+m_D^2f_M(x)}
=2\int_0^1d\xi\xi\frac{1+\xi^2}{4\mu^2\xi^2+m_D^2f_M(\frac{\hat\omega}
{2\xi})}
\eeq
and
\beq
\label{k2e}
\tilde\gamma^{(E)}\simeq \frac{1}{4}\int_{-1}^1d\cos\theta
\frac{1+\cos\theta}{2\mu^2(1-\cos\theta)+m_D^2f_E(x)}
=2\int_0^1d\xi\xi\frac{1-\xi^2}{4\mu^2\xi^2+m_D^2f_E(\frac{\hat\omega}
{2\xi})},
\eeq
where $\hat\omega=\omega/\mu$ and we have changed the
integration variable from $\cos\theta$ to $\xi=\sin\frac{\theta}{2}$.
We first note that the $\xi^2$ terms in the numerator are free from the
collinear singularity.  For these parts the self-energy functions in
the denominators can be dropped and we obtain by trivial integration
a constant $\frac{1}{4\mu^2}$ from ${\cal K}_2^{(M)}$ and a constant
$-\frac{1}{4\mu^2}$ from ${\cal K}_2^{(E)}$.  Amazingly, these
sub-leading contributions of the electric and magnetic parts exactly
cancel in (\ref {k2}). As we shall in section V, this is not the case with
$J\neq 0$.

We now focus on the more difficult infrared sensitive parts of the above
integrals, namely
\beq
\label{cm}
I_M=\int_0^1d\xi\frac{\xi}{4\mu^2\xi^2+m_D^2f_M(\frac{\hat\omega}{2\xi})},
\eeq
and,
\beq
\label{ce}
I_E=\int_0^1d\xi\frac{\xi}{4\mu^2\xi^2+m_D^2f_E(\frac{\hat\omega}{2\xi})}.
\eeq
For the first integral, we introduce $\xi_0$ such that $\hat\omega\ll\xi_0
\ll(\eta\hat\omega)^{\frac{1}{3}}$ with $\eta=\frac{\pi m_D^2}{32\mu^2}$
and divide the integration domain $0<\xi<1$ into the two regions;
$0<\xi<\xi_0$ and $\xi_0<\xi<1$. Correspondingly, we have
$I_M=I_M^>+I_M^<$ with,
\beq
\label{cmout}
I_M^>=\int_{\xi_0}^1d\xi\frac{\xi}{4\mu^2\xi^2+m_D^2f_M
(\frac{\hat\omega}{2\xi})},
\eeq
\beq
\label{cmin}
I_M^<=\int_0^{\xi_0}d\xi\frac{\xi}{4\mu^2\xi^2+m_D^2f_M
(\frac{\hat\omega}{2\xi})}.
\eeq
The Landau damping approximation can then be applied to $\sigma^{M}$ in
(\ref{cmout}) and the integration gives rise to the leading order
logarithm included in $K^\circ$ without any additional constant.  As for
(\ref{cmin}), on the other hand, the mean-value theorem implies that,
\beq
I_M^<=\xi_0\frac{\bar\xi}{4\mu^2\bar\xi^2+m_D^2f_M(\frac{\hat\omega}
{2\bar\xi\vphantom{\vrule height8pt}})},
\eeq
for some $\bar\xi$ in the region $0<\bar\xi<\xi_0$. Since
$f_M(x)$ is a monotonically increasing function of $x$, we find,
\beq
I_M^<\leq\xi_0\frac{\bar\xi}{m_D^2f_M(\frac{\hat\omega}
{2\bar\xi\vphantom{\vrule height8pt}})}
\leq\frac{\xi_0^2}{m_D^2f_M(\frac{\hat\omega}
{2\xi_0})}\simeq \frac{\xi_0^3}{\eta\hat\omega}\ll1.
\eeq
Therefore no additional constant is found from $I_M$ either.

The integral $I_E$ is more subtle than $I_M$ because of the dielectric
behavior at high $\omega$. In this case we have to introduce two cutoffs,
$\xi_0$ and $\xi_0^\prime$ such that $\hat\omega\xi_0\ll\xi_0^\prime
\ll\hat\omega\ll\xi_0\ll1$, and write $I_E=I_E^{(a)}+I_E^{(b)}+I_E^{(c)}$,
where,
\beq
I_E^{(a)}=\int_{\xi_0}^1d\xi\frac{\xi}{4\mu^2\xi^2+m_D^2f_E
(\frac{\hat\omega}{2\xi})},
\eeq
\beq
I_E^{(b)}=\int_{\xi_0^\prime}^{\xi_0}d\xi\frac{\xi}{4\mu^2\xi^2+m_D^2f_E
(\frac{\hat\omega}{2\xi})},
\eeq
and
\beq
I_E^{(c)}=\int_0^{\xi_0^\prime}d\xi\frac{\xi}{4\mu^2\xi^2+m_D^2f_E
(\frac{\hat\omega}{2\xi})}.
\eeq
The integral $I_E^{(a)}$ gives rise to the logarithm included in $K^\circ$
without any additional constant. The mean-value theorem,
applied to $I_E^{(b)}$ implies, with $\xi_0^\prime<\bar\xi<\xi_0$, that,
\beq
I_E^{(b)}=\frac{(\xi_0-\xi_0^\prime)\bar\xi}{4\mu^2\bar\xi^2+m_D^2f_E
(\frac{\hat\omega}{2\bar\xi\vphantom{\vrule height8pt}})}
\leq \frac{\xi_0^2}{m_D^2f_E (\frac{\hat\omega}{2\xi_0'})}
\simeq \frac{\xi_0^2\hat\omega^2}{\xi_0^{\prime2}} \ll1.
\eeq
Finally, the integral $I_E^{(c)}$ is divergent logarithmically at the
lower limit.  However this is an artifact of setting $p=p^\prime=\mu$.
Without this approximation,
one has to consider the partial wave integration simultaneously with the
integrations of $p$ and $p^\prime$. The net result, with $p-p^\prime\sim
\omega$, amounts to a cutoff $\xi\sim\hat\omega^4$ and we expect,
\beq
\label{eq:pcutr}
I_E^{(c)}\sim \frac{1}{4\mu^2}\hat\omega^2\log\frac{1}{\hat\omega},
\eeq
which is completely infrared safe.  Thus we have demonstrated that
there is no constant from ${\cal K}_2^{(M)}+{\cal K}_2^{(E)}$
in addition to the leading logarithms in the limit $\omega\to 0$.  As
the result the first order perturbation (\ref{k2}) leads to no
additional linear log contributions.

\subsection{Second order perturbation.}
Now we proceed to the last three terms of (\ref{pertseries}), which
stand for the second order corrections in ${\cal K}_1+{\cal K}_2$.
Firstly, the penultimate term of (\ref{pertseries}) is easy to dispose
of; although $\langle\tilde 0|({\cal K}_1+{\cal K}_2)|\tilde 0\rangle^2
\sim\log^2\frac{2}{\hat\epsilon}$ this is suppressed by the factor
${\widetilde\lambda_0^{\circ\,^2}}\sim(\log{2\over\hat\epsilon})^{-2}$,
so that it would contribute at most at the constant level, and not at the
linear log level of interest.

Secondly we consider the final term of
(\ref{pertseries}), containing the sum over discrete intermediate
states.  For this term, the arguments given above at first order
also enable us the disregard ${\cal K}_2$ when it is sandwiched
between $|\tilde 0\rangle$ and $|\tilde l\rangle$.  As a result, we only
need to consider $\langle\tilde 0|{\cal K}_1|\tilde l\rangle$ which is at
most linear in logarithms.  Therefore the same power counting argument
used for the penultimate term also applies here.  There is no linear
logarithm associated with the last term and it can be dropped.

Finally, the remaining term involves the second order matrix element
\beq
\langle\tilde0|({\mathcal K}_1+{\mathcal K}_2)^2|\tilde 0\rangle
=I_{\rm{in}}+I_{\rm{out}},
\eeq
where,
\beqa
I_{\rm{in}} &\simeq&
\frac{1}{\beta^3}\sum_{n,m,n^\prime\atop\nu_m<\delta}
\chi_0(\nu_n)\chi_0(\nu_{n^\prime})\int_0^\infty dp\int_0^\infty dq
\int_0^\infty dp^\prime \\ \nonumber
&&\times \frac{[{\cal K}_1(n,p|m,q)+{\cal K}_2(n,p|m,q)]
[{\cal K}_1(m,q|n^\prime,p^\prime)+{\cal K}_2(m,q|n^\prime,p^\prime)]}
{\sqrt{[\nu_n^2+(p-\mu)^2][\nu_{n^\prime}^2+(p^\prime-\mu)^2]}},
\eeqa
and
\beqa
\label{iout}
I_{\rm{out}} &\simeq&
\frac{1}{\beta^3}\sum_{n,m,n^\prime\atop\nu_m>\delta}
\chi_0(\nu_n)\chi_0(\nu_{n^\prime})\int_0^\infty dp\int_0^\infty dq
\int_0^\infty dp^\prime \\ \nonumber
&&\times
\frac{[\tilde\gamma^{J=0}(n,p|m,q)+\tilde\gamma^{J=0}(n,p|-m,q)]
[\tilde\gamma^{J=0}(m,q|n^\prime,p^\prime)
+\tilde\gamma^{J=0}(-m,q|n^\prime,p^\prime)]}
{[\nu_n^2+(p-\mu)^2][\nu_m^2+(q-\mu)^2][\nu_{n^\prime}^2+(p^\prime-\mu)^2]}
\\ \nonumber
&\simeq& \frac{1}{4}\int_{\frac{\hat\epsilon}{2}}^{\hat\delta}
\frac{d\hat\nu}{\hat\nu}
\int_{\frac{\hat\epsilon}{2}}^{\hat\delta}
\frac{d\hat\nu^\prime}{\hat\nu^\prime}
\chi_0(\nu)\chi_0(\nu^\prime)\int_{\delta}^\infty d\nu^{\prime\prime}
\int_0^\infty dq \\ \nonumber
&&\times \frac{\Big[\log\frac{1}{|\nu-\nu^{\prime\prime}|}
+\log\frac{1}{\nu+\nu^{\prime\prime}}\Big]
\Big[\log\frac{1}{|\nu^{\prime\prime}-\nu^\prime|}+
\log\frac{1}{\nu^{\prime\prime}+\nu^\prime}\Big]}
{\nu^{\prime\prime 2}+(q-\mu)^2}.
\eeqa
Consider $I_{\rm{in}}$ first. The leading logarithm of the magnetic
gluon propagator has been subtracted in ${\cal K}_1$ and we may regard
${\cal K}_1$ as bounded in magnitude. The most dangerous contribution comes
from the pole term of the integrations over $p$, $p^\prime$ and $q$. The
summations over $n$, $m$ and $n^\prime$ give rise to a contribution of order
at most $\log^2\frac{2}{\hat\epsilon}$ (taking into account the inverse
square root of $\log\frac{2}{\hat\epsilon}$ from the normalization constant
of each $\chi_0$).  The net power of logarithms, when multiplied by
$\widetilde\lambda^{\circ\,^2}$ is thus zero. As to $I_{\rm{out}}$,
since the intermediate energy $\nu^{\prime\prime}$ is kept away from the
Fermi level, the logarithms in the numerator of (\ref {iout}) are smeared.
We are left with only two powers of logarithms from the lower limits of
the $\nu$ and $\nu^\prime$ integrations and one inverse power of logarithm
from the normalization constant of $\chi_0(\nu)$. The net power of logarithms,
when multiplied by $\widetilde\lambda^{\circ\,^2}$, is $-1$.
Therefore we do not find any
contribution to the linear logarithmic term from second order
perturbation theory.

In fact the only non-vanishing contribution at the
linear log level, (\ref{cont-disc}), is due to the discrete to continuous
energy correction, and is obtained at first order in the perturbative
expansion. Putting these results together, the expansion of the inverse
square of the largest eigenvalue associated to the Fredholm equation with
free quark propagators reads
\beq
\label{res}
\frac{1}{\widetilde\lambda_0^2}=\frac{1}{\widetilde\lambda_0^{\circ\,^2}}
+{8\over\pi^2}(\gamma+\log2)\log\frac{2}{\hat\epsilon}+{\cal O}(1).
\eeq
The linear logarithmic term of (\ref {res}) contributes to the factor
$c_1^{\prime\prime}$ of (\ref{eq:c1pp}).
Our analytical arguments on the perturbation corrections are confirmed by
numerical integration, as will be discussed in section VI.
%

\section{Non-zero Angular Momentum.}
Although we have focused on the $s$-wave channel for longitudinal
pairs of quarks, the collinear singularity of one-gluon exchange which
is responsible for the long-range nature of the pairing interaction
makes no distinction at leading order between zero and non-zero
angular momenta.  Therefore, in any angular momentum $J$ channel, the
largest factor contributing to the critical temperature, $k_BT_C^J$,
is given by the non-BCS relation (\ref{sonsfactor}) in all cases.
However, the constant pertaining to the logarithm may vary with
angular momentum.  For the $J^{\rm th}$ partial wave, the increasing
number of nodes of the Legendre polynomial $P_J(\cos\theta)$ with
increasing $J$ reduces the effective attraction between quarks.  As we
shall see, although formally of ${\mathcal O}(1)$, the suppression for
$J \neq 0$ can be of several orders of magnitude.

To explore such a correlation, we evaluate the $J\neq 0$
partial wave component of the irreducible vertex function,
\beqa
\label{jfull}
\tilde\gamma^J(n^\prime,p^\prime|n,p) &=& -\frac{g^2}{8}\left(1+
\frac{1}{N}\right) (2J+1) \int_{-1}^{1}
d(\cos \theta) P_J(\cos\theta)\\ \nonumber
&&\times\Big[\frac{3 - \cos \theta
- (1 + \cos \theta) \frac{(p - p')^2}{ |\vec{p} - \vec{p}\,'|^2}}
{(\nu_n - \nu_{n'})^2 + |\vec{p} - \vec{p}\,'|^2 + m_D^2f_M(x)}
+ \frac{1 + \cos \theta}{|\vec{p} - \vec{p}\,'|^2 + m_D^2f_E(x)}\Big].
\eeqa
To uncover the dependence on $J$, we set $p=p'=\mu$. The numerator of
the integrand can be decomposed into a constant term and a non-constant
term with the latter vanishing in the forward direction,
$\cos\theta=1$.  Since $P_J(1)=1$ for any $J$, the contribution from the
constant term is independent of $J$ and can be approximated by the $J=0$
vertex given by Eqn.~(\ref{approx}).  For the non-constant part of the
numerator, proportional to $1-\cos\theta$, the forward direction is
infrared safe, and the self energies $f_M$ and $f_E$ may be
dropped to the order we are interested in. The resultant integral produces
a sub-leading $J$-dependent term.  We find,
\begin{equation}
\label{pj}
\tilde\gamma^J(n',p'|n,p) = -\frac{g^2}{12\mu^2} \left(
1 + \frac{1}{N}\right) (2J+1)
\Big(\log \frac{1}{|\hat\nu_n-\hat\nu_{n^\prime}|}+3c_J\Big),
\end{equation}
with,
\begin{equation}
\label{const}
c_J=\int_{-1}^{1}d\cos\theta\frac{P_J(\cos\theta)-1}{1-\cos\theta}.
\end{equation}

Using the recursion formula for Legendre polynomials,
\begin{equation}
(J+1)P_{J+1}(\cos\theta)-(2J+1)\cos\theta P_J(\cos\theta)
+JP_{J-1}(\cos\theta)=0,
\end{equation}
we find that,
\begin{equation}
\label{cj}
(J+1)c_{J+1}-(2J+1)c_J+Jc_{J-1}=-2\delta_{J\,0}.
\end{equation}
Since $c_0=0$, the solution to (\ref{cj}) for $J \geq 1$ is,
\begin{equation}
\label{const2}
c_J=-2\sum_{n=1}^J\frac{1}{n}.
\end{equation}
Following the arguments of the perturbation theory for $J=0$, we see
that $c_J$ contributes only to the first order perturbation of the
eigenvalue, and hence the critical temperature in the angular momentum $J$
channel reads,
\begin{eqnarray}
\label{scl}
k_BT_C^{J\neq 0} &=& e^{3 c_J} k_B T_C^{J=0} \\
&\simeq&2.5 \times 10^{-3}~ k_B T_C^{J=0} \qquad\hbox{for $J=1$},\nonumber
\end{eqnarray}
where $T_C^{J=0}$ is the corresponding critical temperature in the
$J=0$ channel.

The foregoing consideration can be generalized to transverse pairings
between opposite helicities. The only modifications include replacing
$3-\cos\theta$ in (\ref{jfull}) by $1+\cos\theta$ and $P_J(\cos\theta)$ by
$d^J_{11}(\theta)$, the Wigner D-functions \cite{goldberg}. The final
formula for the transition temperature is given by (\ref{scl}) with
$c_J$ replaced by
\beq
\frac{1}{2}\Big(c_J+\frac{J}{2J+1}c_{J+1}+\frac{J+1}{2J+1}c_{J-1}\Big),
\eeq
with $J\ge 1$ and $c_0=0$.  Relative to the longitudinal pairing at $J=0$,
we find that for $J=1$ a longitudinal pairing is suppressed by a factor
of $e^{-6}$ whereas a transverse pairing is suppressed by $e^{-4.5}$.

Physically, (\ref{scl}) shows a large suppression of the transition
temperature with non-zero angular momentum relative to that for
$s$-wave pairing.  As the quark-gluon plasma is cooled from the normal phase,
the first critical temperature reached is $T_C^{J=0}$ and the
corresponding long range order is then switched on. This would then
invalidate the relation (\ref{scl}) for $J\neq 0$, which was derived
using normal phase propagators. Furthermore, there is no simple relation
between $T_C^{J\neq 0}$ and the corresponding gap energy at $T=0$.
Nevertheless, the suppression in (\ref{scl}) indicates that if for some
physical reasons the long range order with $J= 0$ cannot be formed, then
the $J\neq 0$ pairing can only happen at very low temperatures.
%

\section{Numerical Results and Gauge Invariance}

In the previous section we have examined the perturbative expansion of the
full eigenvalue problem, (\ref{326}), and have found a single correction
term at the linear log level, arising from the discrete to continuum energy
approximation, (\ref{cont-disc}).  While this result is simply stated, it
came from an involved perturbative treatment of the Fredholm integral
equation to second order.  Thus, as a check, the analytical arguments on
the perturbation corrections have been confirmed by numerical evaluation
of both the first and second order terms in the perturbative expansion,
(\ref{pertseries}).

At this point, a comment on the integration limits is in order.  We
recall that a cutoff, $\hat\nu < \hat\delta$,  was imposed in order to
control the zeroth order kernel, (\ref{dl}).  As this cutoff is not
otherwise physical, its dependence ought to drop out of the final
result.  This in fact has been demonstrated in the toy model of section
II, where the cutoff dependence cancels between the zeroth and first
order terms.  For the actual case at hand, however, the cutoff
dependence behaves differently.  Consider the zeroth order one-gluon
exchange kernel, (\ref{sons}), which may be rewritten in terms of the
variable $x=\log\hat\nu$:
\begin{equation}
\label{eq:sonab}
f(x)=-\widetilde\lambda^2\int_a^b dx' x_> f(x'),
\end{equation}
where $a=\log{\hat\epsilon\over2}$ and $b=\log\hat\delta$.  The exact
eigenvalues of (\ref{eq:sonab}) may be obtained using Son's method
\cite{son1998}.  The resulting spectrum is given by the zeros of the
transcendental equation,
\begin{equation}
\label{eq:stran}
b\widetilde\lambda=-\cot\widetilde\lambda(b-a),
\end{equation}
and may be developed perturbatively for $(b/a)\ll1$.  We find,
\begin{equation}
\label{eq:stranpert}
{1\over\widetilde\lambda_j^2} =
[(j+{\textstyle{1\over2}})\pi]^{-2}\left(\log{2\over\hat\epsilon}\right)^2
\left[1+{2\over3}[(j+{\textstyle{1\over2}})\pi]^2
\left({\log\hat\delta\over\log{2\over\hat\epsilon}}\right)^3
+ O \left({\log\hat\delta\over\log{2\over\hat\epsilon}}\right)^5\right],
\end{equation}
which corresponds to Eqn.~(\ref{eq:cevals}) for $\hat\delta\ne1$.  This
indicates that the cutoff dependence of the zeroth order term is in fact
down by three
powers of the logarithm, and hence contributes below the level we are
interested in.  This in fact explains why, unlike the toy model, we find
no cutoff dependence at the linear log level term by term in the
perturbative expansion (\ref{pertseries}).  Thus our numerical work for the
above was performed with a fixed upper cutoff of $\hat\delta=1$.

In addition to numerical confirmation of the perturbative expansion, we
have also investigated the numerical solution to the Fredholm equation
(\ref{326}).  This provides both a direct verification of our analytic
results and a preliminary demonstration of the gauge invariance of our
approach.  Working in the normal phase makes treatment of gauge invariance
easier to handle.  In particular, as discussed in \cite{BLR99}, the
Fredholm determinant can be represented as a series of bubble diagrams,
the sum of which is gauge invariant.

As this numerical work is secondary to the analytical results presented
above, we make several simplifying approximations.  Firstly, to avoid
infinite sums, we only work with a continuous energy integral.  As a
consequence, we will not reproduce the discrete to continuum energy
correction of Eqn.~(\ref{cont-disc}).  Secondly, we {\it a priori}
dispose of the momentum integral by noting that the
dominant contribution to the kernel occurs near the Fermi surface,
$\{p,p'\}\approx\mu$.  These two approximations correspond to the steps
taken in Eqns.~(\ref{eq:matsum}) and (\ref{eq:pint}) in deriving the
zeroth order integral equation, (\ref{sons}).  Numerically, however,
we do not employ Son's approximation, (\ref{sonsapprox}), and instead
use the complete one gluon
exchange vertex (evaluated with momenta at the Fermi surface).

One difficulty immediately arises as a consequence of setting $p=p'=\mu$
in the one gluon exchange kernel, and that is the spurious divergence in
the Coulomb propagator, (\ref{eq:glue4}), in the forward direction.  While
this divergence is eliminated by a full treatment incorporating momenta off
of the Fermi surface, as shown in (\ref{eq:pcutr}), here we have no such
luxury without reintroducing the momentum integral.  This difficulty prompts
us to work in the covariant gauge, using the propagators of \cite{SWc}.
Using the covariant gauge propagators directly in numerical analysis of the
integral equation provides a first check upon the gauge invariance of the
approach.

\begin{figure}[t]
\epsfxsize 10cm
\centerline{\epsffile{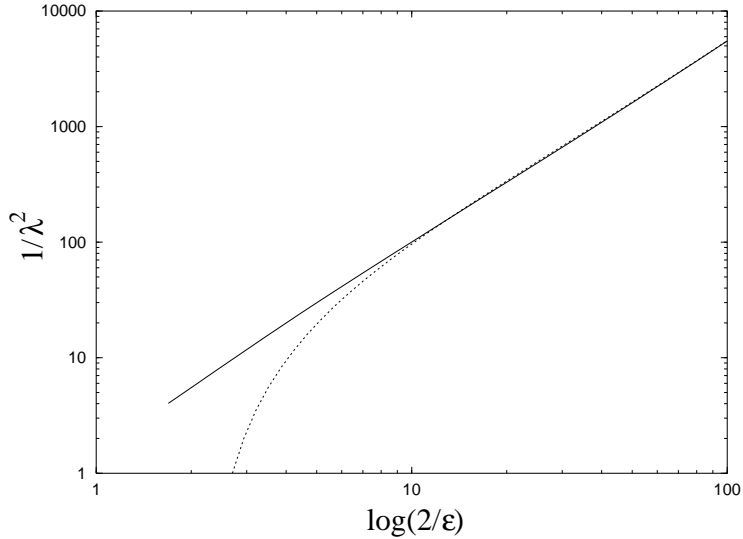}}
\bigskip
\caption{The largest eigenvalue, $1/\widetilde\lambda^2$, of the
Fredholm problem (\ref{eq:numfred}) as a function of $\log{2\over\epsilon}$.
The numerical solution is given by the solid line, while the zeroth order
solution, (\ref{eq:anazeval}), is given by the dashed line.}
\label{fig:fred1}
\end{figure}

Taking everything into consideration, we thus examine the numerical
solution to the integral equation,
\begin{equation}
\label{eq:numfred}
\chi(\nu) = \widetilde\lambda^2\int_{\epsilon\over2}^{\delta}
{d\nu'\over\nu'} \widetilde K(\nu,\nu')\chi(\nu'),
\end{equation}
with kernel
\begin{equation}
\widetilde K(\nu,\nu')=3[\gamma(\nu,\nu')+\gamma(\nu,-\nu')].
\end{equation}
For the one gluon exchange interaction in covariant gauge, we use the
result of \cite{SWc}.  Namely, in our notation,
\begin{equation}
\label{eq:swccov}
\gamma(\nu,\nu') = 2\int_0^1\!\!d\xi\xi\!\left[
{1+\xi^2\over(\nu-\nu')^2+4\mu^2\xi^2+m_D^2f_M(x)}
+{1-\xi^2\over(\nu-\nu')^2+4\mu^2\xi^2+m_D^2(1+x^2)f_E(x)}
\right],
\end{equation}
where $x=|\nu-\nu'|/(2\mu\xi)$.  Note that the electric gluon exchange
term is modified in covariant gauge, as may be seen by comparison with
(\ref{k2m}) and (\ref{k2e}).  For this linear Fredholm problem, which is
similar to the non-linear gap equation, the numerical solution for the
largest eigenvalue is easily obtained by iteration.  One subtlety arises
in this case, however, in that the relation between $T_C$ and the gauge
coupling $g$ is given by inversion of the critical temperature
condition, $\lambda^2(T_c,g,\mu)=1$.  To investigate the behavior of
(\ref{eq:numfred}) near the critical point, we thus use the zeroth order
relation, $g^2(1+{1\over N})=6\pi^4(\log{2\over\epsilon})^{-2}$, where
$\epsilon = 2\pi k_BT_C$, to fix $m_D^2$ in (\ref{eq:swccov}).  For
explicit numerical work, we furthermore set $\delta=10^{10}\mu$, thus
employing a finite but large cutoff.

\begin{figure}[t]
\epsfxsize 10cm
\centerline{\epsffile{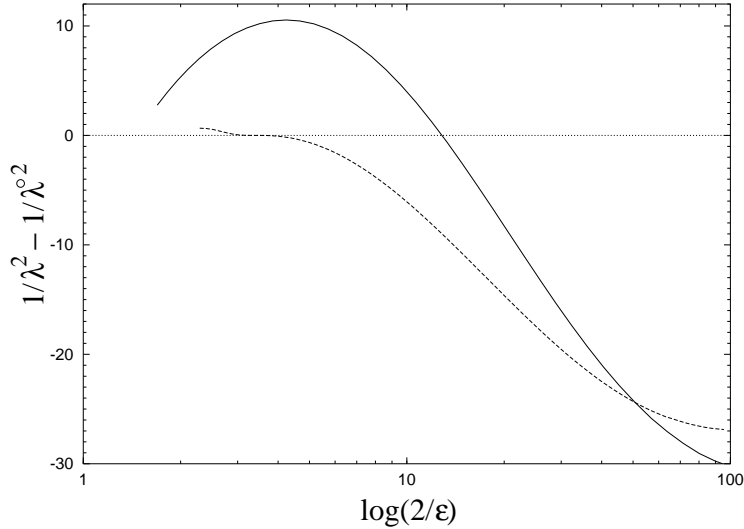}}
\bigskip
\caption{The difference between the numerical solution and the analytic
result, ${1\over\widetilde\lambda^{\circ\,^2}} =
{4\over\pi^2}\log^2{2\over\hat\epsilon}$ (solid line).  For comparison,
the $\hat\delta$-dependent behavior of the eigenvalue, given by
(\ref{eq:stran}), is indicated by the dashed line.  Note the different
scale used on the vertical axis, compared with Fig.~\ref{fig:fred1}.}
\label{fig:fred2}
\end{figure}

The numerical solution to (\ref{eq:numfred}) is presented in
Fig.~\ref{fig:fred1}, and shows good agreement with the analytic result,
\begin{equation}
\label{eq:anazeval}
{1\over\widetilde\lambda^{\circ\,^2}}={4\over\pi^2}\log^2{2\over\hat\epsilon}
= {4\over\pi^2}
\left(\log{2\over\epsilon} + \log{256\mu^6\over\pi m_D^5} \right)^2,
\end{equation}
at least for $\log{2\over\epsilon}\gg1$, corresponding to $g\ll1$.  This
behavior is highlighted in Fig.~\ref{fig:fred2}, where the difference
between the numerical and analytical solutions is shown.  Up to the
linear log level, there is no discrepancy between the solutions; as they
were obtained in different gauges, these results indeed support the gauge
invariance of the critical temperature.

The behavior in Fig.~\ref{fig:fred2} clearly indicates the presence
of subleading terms, contributing below the linear log level.  A large
portion of this appears to be related to the treatment of the
ultraviolet cutoff, taken as $\hat\delta=1$ in the analytical work.  The
numerical work, on the other hand, is suggestive of the presence of an
effective cutoff near the Fermi surface, $\delta\approx\mu$, corresponding
to $\hat\delta=(\pi/256)(m_D/\mu)^5$.
While any difference between these cases
is formally seen from (\ref{eq:stranpert}) to be down by
three powers of the logarithm, it however contains a hidden
enhancement due to the factor of $m_D^5$ entering into $\hat\delta$.
Since $m_D\sim g\sim (\log{2\over\epsilon})^{-1}$ (using the zeroth
order relation), the relevant expansion parameter is
$(\log\hat\delta/\log{2\over\hat\epsilon})\sim (5\log x)/x$ where
$x = \log{2\over\epsilon}$.  Even for $x\approx100$, which is the largest
value we examined numerically, its cube is only down by about one
percent.

For relevant values of $g$, which are of order one, the values of $x$
range from about 5 to 30, in which case this sensitivity to the cutoff
can lead to effects of order 10\%.  On the other hand, for the nonlinear
gap equation, this sensitivity appears much more pronounced; varying the
ultraviolet cutoff between $\delta$ and $\mu\hat\delta$ can lead to as
much as a factor of two or more change in the value of the zero
temperature gap!

\section{Conclusion.}
In this paper, we present an exact calculation of the transition
temperature to the color-superconducting phase of $SU(N)_c$ QCD in the
limit of high baryon density.  To do this we cast the Fredholm
integral equation that arises from the Dyson-Schwinger approach to
di-quark scattering into a general eigenvalue problem.  The leading
order of the eigenvalue, found to be doubly logarithmic, gives rise to
the non-BCS dependence of the exponent upon the coupling.  We have
developed a perturbation method which enables us to calculate the
eigenvalue beyond the leading order in logarithm.  The result,
together with the contribution from the radiative correction of the
quark propagator found in \cite{BLR99}, leads to a complete
determination of both the exponential and the pre-factor of the
scaling formula for the transition temperature, (\ref{TC}), to the
leading order of the running coupling constant.  The transition
temperature and the zero temperature gap are both suppressed relative
to previous claims due to the effects of the quark self-energy.  For
$N=N_f=3$, these effects reduce the results by a significant factor
of $0.17663$.

\begin{figure}[t]
\epsfxsize 10cm
\centerline{\epsffile{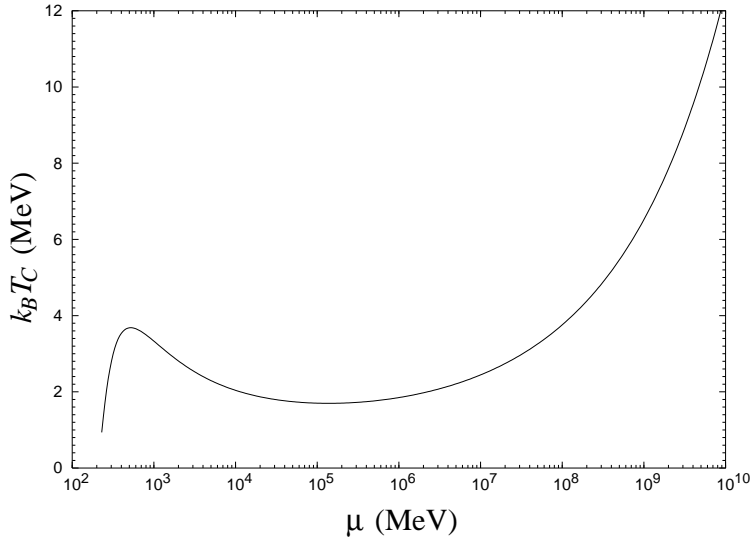}}
\bigskip
\caption{The transition temperature, $k_BT_C$, as a function of the
chemical potential, $\mu$, for $N=N_f=3$ and
$\Lambda = 200 {\rm MeV}$.}
\label{fig3}
\end{figure}

Using the one-loop formula for the running coupling constant at the
chemical potential,
\beq
\label{run}
g^2=\frac{12\pi^2}{\Big(\frac{11}{2}N-N_f\Big)\log \frac{\mu}{\Lambda}},
\eeq
the behavior of the transition temperature is shown in Fig.~\ref{fig3}
for a longitudinal pairing with zero angular momentum for
$N=N_f=3$ and $\Lambda=200{\rm{MeV}}$.  We note that $g\simeq 1$
corresponds to $\mu \simeq 10^6 {\rm MeV}$ and a transition
temperature $k_BT_C \simeq 1.9 {\rm MeV}$.  Our calculations become valid
in the limit of high baryon density where the theory is weakly
coupled, corresponding to the upward sloping arm on the right of
Fig.~\ref{fig3} where $\mu > 10^6{\rm MeV}$.  However, there is
some contention that such
perturbative series are valid for $g \leq 4$, see \cite{pisarski1999c}
and the references therein.  Extrapolating our results to $g \sim 3$
corresponds to $\mu \simeq 500{\rm MeV}$ and $k_BT_C \simeq 3.7
{\rm MeV}$.  This is the point at which the graph in Fig.~\ref{fig3}
turns over and the behaviour becomes unphysical.  This
limit suggests that for $g > 3$ strong coupling effects truly dominate and
the transition temperature is determined by the mechanisms of chiral
symmetry breaking and confinement but for $g < 3$ our results may prove
increasingly accurate.

The ratio between the transition temperature and the energy gap at
zero temperature, $\Delta$, is an important number for
superconductivity.  This ratio has been calculated for zero angular
momentum in \cite{pisarski1999c} and the result is identical to that
of BCS theory, {\it i.e.}
\beq
\label{bcs}
\frac{\Delta}{k_BT_C}=\pi e^{-\gamma}.
\eeq
Combining our formula for $k_BT_C^{J=0}$, (\ref{TC}), and this ratio we
obtain that,
\begin{equation}
\label{TC2}
\Delta=2c_1'\Delta_0
=512 c_1^\prime\pi^4 \left( \frac{2}{N_f} \right)^{\frac{5}{2}}
\frac{\mu}{g^5}
e^{-\sqrt{6N\over N+1}{\pi^2\over g}},
\end{equation}
with $c_1^\prime$ given by (\ref{c1}) and $\Delta_0$ given by
(\ref{sonsfactor}) and previously
determined in \cite{SWc}.  Apart from the new factor
$c_1'$, this is in agreement with the gap energy presented in
\cite{pisarski1999c} (we have highlighted the factor of 2 which was
found numerically in \cite{SWc} and by analytic means
in \cite{pisarski1999c}).  What is significant, and perhaps somewhat
surprising, is that the additional unknown pre-factor of ${\mathcal
O}(1)$ present in all previous calculations of the zero temperature
gap has been found to be exactly one to the leading order in the coupling.

We have extended the analysis to find the transition temperature for
both longitudinal and transverse pairings for arbitrary angular
momentum, $J\geq 1$.  Relative to the longitudinal $J=0$ pairing these
channels are suppressed by several orders of magnitude.  As the
temperature is reduced, once the highest transition temperature is
reached the results for lower transition temperatures are invalid.
However, if for some physical reason the $J=0$ longitudinal pair
cannot form these results indicate that color-superconductivity may
only occur at very low temperatures.  Further investigation of the
ratio (\ref{bcs}) generalised to $J >0$ may produce interesting
results for condensates of non-zero angular momentum.

\bigskip

\noindent
{\bf Acknowledgements.}

W.B.~and H-C.R.~would like to thank R.~Pisarski and D.~Rischke for
stimulating discussions.  The work of W.B.~and H-C.R.~is supported
in part by the U.S.~Department of Energy under grant DOE-91ER40651-TASKB.
H-C.~Ren's work is also supported in part by the Wiessman visiting
professorship of Baruch College of CUNY.

\begin{appendix}

\section{Single Gluon Exchange Vertex.}

In this appendix, we set out some conventions and tabulate the single-gluon
exchange vertex in all channels with $s=\pm$ and $s^\prime=\pm$. In
Coulomb gauge, the relevant one gluon exchange contribution to
(\ref{eq:symanti}) reads
\beqa
%
%
\Gamma^A_{s^\prime,s}(n^\prime,\vec p\,^\prime |n,\vec p\,) &=&
-i\frac{g^2}{2}\Big(1+\frac{1}{N}\Big)D^M(\vec k,\omega)
[\bar u(s_1^\prime\vec p\,^\prime)\gamma_iu(s_1^{\vphantom{\prime}}\vec p\,)]
[\bar u(-s_2^\prime\vec p\,^\prime)\gamma_j
u(-s_2^{\vphantom{\prime}}\vec p\,)]
(\delta_{ij}-\hat k_i\hat k_j)\nonumber\\
&&- i\frac{g^2}{2}\Big(1+\frac{1}{N}\Big)D^E(\vec k,\omega)
[u^\dagger(s_1^\prime\vec p\,^\prime)u(s_1^{\vphantom{\prime}}\vec p\,)]
[u^\dagger(-s_2^\prime\vec p\,^\prime)u(-s_2^{\vphantom{\prime}}\vec p\,)],
\eeqa
where $\vec k=\vec p\,^\prime-\vec p$ and $\omega=\nu^\prime-\nu$.
Using the Dirac equation $(-i\vec\gamma\cdot\vec p+\gamma_4p)u(\vec p)=0$
and introducing the notation
\beq
%
%
V_{s_1^\prime,s_2^\prime;s_1^{\vphantom{\prime}},s_2^{\vphantom{\prime}}}=
[\bar u(s_1^\prime\vec p\,^\prime)\vec\gamma
u(s_1^{\vphantom{\prime}}\vec p\,)]
[\bar u(-s_2^\prime\vec p\,^\prime)\vec\gamma
u(-s_2^{\vphantom{\prime}}\vec p\,)]
\eeq
and
\beq
S_{s_1^\prime,s_2^\prime;s_1^{\vphantom{\prime}},s_2^{\vphantom{\prime}}}=
[u^\dagger(s_1^\prime\vec p\,^\prime)u(s_1^{\vphantom{\prime}}\vec p\,)]
[u^\dagger(-s_2^\prime\vec p\,^\prime)u(-s_2^{\vphantom{\prime}}\vec p\,)],
\eeq
we have
\beqa
\Gamma_{s^\prime,s}^A(n^\prime,\vec p\,^\prime|n,\vec p)
&=&-i\frac{g^2}{2}\left(1+\frac{1}{N}\right)D^M(\vec k,\omega)\Big[
V_{s_1^\prime,s_2^\prime;s_1^{\vphantom{\prime}},s_2^{\vphantom{\prime}}}
-\frac{(s_1^\prime p^\prime-s_1^{\vphantom{\prime}}p)
(s_2^\prime p^\prime-s_2^{\vphantom{\prime}}p)}{(\vec p\,^\prime-\vec p\,)^2}
S_{s_1^\prime,s_2^\prime;s_1^{\vphantom{\prime}},s_2^{\vphantom{\prime}}}\Big]
\nonumber \\
&&-i\frac{g^2}{2}\left(1+\frac{1}{N}\right)D^E(\vec k,\omega)
S_{s_1^\prime,s_2^\prime;s_1^{\vphantom{\prime}},s_2^{\vphantom{\prime}}}.
\eeqa
Our convention for gamma matrices is that $\{\gamma_\mu, \gamma_\nu\}=
\delta_{\mu\nu}$ with each $\gamma_\mu$ hermitian and
\beq
\gamma_5=\gamma_1\gamma_2\gamma_3\gamma_4=
\left(\matrix{0&-I\cr-I&0\cr}\right).
\eeq
The coordinate system is chosen such that $\vec p\parallel\hat z$ and
$\vec p\,^\prime\perp\hat y$ with $\vec p\,^\prime\cdot\hat x>0$. The
azimuthal angle of $\vec p$ is assigned to be 0 and that of $-\vec p$
to be $\pi$.  The polar angles of $\pm\vec p$ and $\pm\vec p\,^\prime$
are $\vec p:\,(0,0)$, $-\vec p:\,(\pi,\pi)$, $\vec p\,^\prime:\,(\theta,0)$
and $-\vec p\,^\prime:\,(\pi-\theta,\pi)$. The chiral spinors at $\pm\vec p$
and $\pm\vec p\,^\prime$ are related via
\beq
u(\vec q\,)=e^{-\frac{i}{2}\sigma_3\phi}e^{-\frac{i}{2}\sigma_2\theta}
e^{\frac{i}{2}\sigma_3\phi}u(\vec p\,),
\eeq
with $\vec q$ any one of $-\vec p$ and $\pm\vec p\,^\prime$ with
$(\theta,\phi)$ the corresponding polar angles. For $\gamma_5 u(\vec p)=
-u(\vec p)$ we write
\beq
u(\vec p\,)=\frac{1}{\sqrt{2}}\left(\matrix{1\cr 0\cr 1\cr 0\cr}\right).
\eeq
Then the expressions for
$V_{s_1^\prime,s_2^\prime;s_1^{\vphantom{\prime}},s_2^{\vphantom{\prime}}}$
and
$S_{s_1^\prime,s_2^\prime;s_1^{\vphantom{\prime}},s_2^{\vphantom{\prime}}}$
follow and are listed in Table.~I.

\begin{table}
\begin{tabular}{c|r|r}
$s_1's_2',s_1^{\vphantom{\prime}}s_2^{\vphantom{\prime}}$&
$V_{s_1^\prime,s_2^\prime;s_1^{\vphantom{\prime}},s_2^{\vphantom{\prime}}}$
\kern12pt&
$S_{s_1^\prime,s_2^\prime;s_1^{\vphantom{\prime}},s_2^{\vphantom{\prime}}}$
\kern12pt\\
\hline
$++,++$&$\frac{3}{2}-\frac{1}{2}\cos\theta$\kern12pt&
       $\frac{1}{2}+\frac{1}{2}\cos\theta$\kern12pt\\
$+-,++$&$\frac{1}{2}\sin\theta$\kern12pt&
       $-\frac{1}{2}\sin\theta\kern12pt$\\
$-+,++$&$-\frac{1}{2}\sin\theta$\kern12pt&
       $\frac{1}{2}\sin\theta$\kern12pt\\
$--,++$&$-\frac{3}{2}-\frac{1}{2}\cos\theta$\kern12pt&
       $-\frac{1}{2}+\frac{1}{2}\cos\theta$\kern12pt\\
$++,+-$&$-\frac{1}{2}\sin\theta$\kern12pt&
       $\frac{1}{2}\sin\theta$\kern12pt\\
$+-,+-$&$-\frac{1}{2}-\frac{1}{2}\cos\theta$\kern12pt&
       $\frac{1}{2}+\frac{1}{2}\cos\theta$\kern12pt\\
$-+,+-$&$-\frac{1}{2}+\frac{1}{2}\cos\theta$\kern12pt&
       $\frac{1}{2}-\frac{1}{2}\cos\theta$\kern12pt\\
$--,+-$&$\frac{1}{2}\sin\theta$\kern12pt&
       $\frac{1}{2}\sin\theta$\kern12pt\\
$++,-+$&$\frac{1}{2}\sin\theta$\kern12pt&
       $-\frac{1}{2}\sin\theta$\kern12pt\\
$+-,-+$&$-\frac{1}{2}+\frac{1}{2}\cos\theta$\kern12pt&
       $\frac{1}{2}-\frac{1}{2}\cos\theta$\kern12pt\\
$-+,-+$&$-\frac{1}{2}+\frac{1}{2}\cos\theta$\kern12pt&
       $\frac{1}{2}+\frac{1}{2}\cos\theta$\kern12pt\\
$--,-+$&$-\frac{1}{2}\sin\theta$\kern12pt&
       $-\frac{1}{2}\sin\theta$\kern12pt\\
$++,--$&$-\frac{3}{2}-\frac{1}{2}\cos\theta$\kern12pt&
       $-\frac{1}{2}+\frac{1}{2}\cos\theta$\kern12pt\\
$+-,--$&$\frac{1}{2}\sin\theta$\kern12pt&
       $-\frac{1}{2}\sin\theta$\kern12pt\\
$-+,--$&$-\frac{1}{2}\sin\theta$\kern12pt&
       $\frac{1}{2}\sin\theta$\kern12pt\\
$--,--$&$\frac{3}{2}-\frac{1}{2}\cos\theta$\kern12pt&
       $\frac{1}{2}+\frac{1}{2}\cos\theta$\kern12pt
\end{tabular}
\bigskip
\caption{The factors $V$ and $S$ for all combinations of incoming and
outgoing quarks above and below the Fermi Sea.}
\label{tbl:llc}
\end{table}

\section{First Order Corrections From ${\mathcal K}_1^{(b)}$.}

In this appendix, we shall calculate the normalization constant of
$|j\rangle$
due to discrete-continuous energy corrections, and calculate the
contribution of ${\cal K}_1^{(b)}$ giving rise to (\ref{cont-disc}).
We begin with
a simplification of the matrix elements of an operator of the structure
\beq
\langle\nu^\prime,p^\prime|O|\nu,p\rangle=\frac{2o(\nu^\prime,\nu)}
{\sqrt{\nu^2+(p-\mu)^2}\sqrt{\nu^{\prime 2}+(p^\prime-\mu)^2}}
\eeq
For a function $\langle\nu,p|\psi\rangle$ pertaining to the ket vector
$|\psi\rangle$, we introduce the function
\beq
\langle\nu|\psi\rangle=\sqrt{2}\int_0^\infty dp
\frac{\langle\nu,p|\psi\rangle}{\sqrt{\nu^2+(p-\mu)^2}},
\eeq
so that
\beq
\langle\psi^\prime|O|\psi\rangle=\sum_{n=0}^\infty\frac{1}
{\left(n^\prime+\frac{1}{2}\right)\left(n+\frac{1}{2}\right)}
\langle\psi^\prime|\nu_{n^\prime}\rangle o(\nu_{n^\prime},\nu_n)
\langle\nu_n|\psi\rangle.
\eeq
Note that we have extended the lower limit of the $p$ and $p^\prime$
integrals to $-\infty$; the correction introduced is beyond the order of
interest. The same simplification extends to the next appendix as well.

Given ${\cal K}_1^{(b)}$ in Eqn.~(\ref{eq:kern1b}), we wish to evaluate the
contribution
\begin{equation}
\label{eq:k1bpt}
\langle\tilde0|{\cal K}_1^{(b)}|\tilde0\rangle
= {\langle0_c|\log{1\over\hat\nu_>}|0_c\rangle\over\langle0_c|0_c\rangle}
-{1\over\widetilde\lambda_0^{\circ\,^2}}.
\end{equation}
Note that while the eigenfunction,
\begin{equation}
\label{eq:zefun}
\langle\nu|0_c\rangle = \chi_0(\nu) =
{2\sqrt{t\over\pi}} \sin\left(t\log{1\over\hat\nu}\right),
\end{equation}
where
\begin{equation}
t={\pi\over 2\log{2\over\hat\epsilon}}\ll1,
\end{equation}
is properly normalized in the continuum norm, this may no long be the
case under the discrete norm.  Thus we must include the denominator
$\langle0_c|0_c\rangle$ in (\ref{eq:k1bpt}), which is to be calculated with
discrete energy variables.

Using (\ref{eq:zefun}), this discrete norm of the zeroth order
eigenfunction reads,
\begin{equation}
\langle0_c|0_c\rangle={4t\over\pi}\sum_{n=0}^{N_0}{\hat\epsilon\over\hat\nu_n}
\sin^2 \left(t\log{1\over\hat\nu_n}\right),
\end{equation}
where
\begin{equation}
\hat\nu_n=(n+{\textstyle{1\over 2}})\hat\epsilon
\end{equation}
and $N_0\gg1$ is the discrete cutoff corresponding to the ultraviolet
cutoff, $\hat\delta$, in the continuum: $\hat\delta=(N_0+{1\over 2})
\hat\epsilon$.

Evaluation of this sum may be performed in terms of the generalized
zeta-function, \cite{whittaker},
\begin{equation}
\zeta(s,a)=\sum_{n=0}^\infty{1\over (n+a)^s},
\end{equation}
resulting in,
\begin{eqnarray}
\langle0_c|0_c\rangle&=&{2t\over\pi}
\Bigl\{\zeta(1+0^+,{\textstyle{1\over 2}})
-\zeta(1+0^+, N_0+{\textstyle{3\over 2}})\nonumber\\
&&\qquad\qquad
-{\rm{Re}}\, \hat\epsilon^{-2it}\left[\zeta(1+2it,{\textstyle{1\over 2}})
-\zeta(1+2it, N_0+{\textstyle{3\over 2}})\right]\Bigr\}.
\end{eqnarray}
Using the asymptotic formula for $t\to 0$,
\begin{equation}
\zeta(1+2it,a)=-i\frac{\Gamma(a)}{2t\Gamma(a+it)}+{\cal O}(t),
\end{equation}
which follows the integral representation of $\zeta(s,a)$, we obtain that
\beq
\zeta(1+2it,\frac{1}{2})\simeq -i\frac{1}{2t}-\psi\left(\frac{1}{2}\right),
\eeq
with $\psi(z)$ the logarithmic derivative of $\Gamma(z)$ and
\begin{equation}
\zeta(1+2it,N_0+\frac{3}{2})\simeq-\frac{i}{2t}
\Big(N_0+\frac{3}{2}\Big)^{-2it}.
\end{equation}
Therefore, we end up with
\begin{equation}
\label{eq:disnorm}
\langle0_c|0_c\rangle \simeq 1-{4t\over\pi}(\psi({\textstyle{1\over2}})+\log2)
=1+2(\gamma+\log2)\log^{-1}{2\over\hat\epsilon},
\end{equation}
where we have used $\psi({1\over2})=-\gamma-2\log2$.

Evaluating the matrix element of the kernel is more difficult.  We start
with the expression
\begin{equation}
\langle0_c|\log{1\over\hat\nu_>}|0_c\rangle
={4t\over\pi}\sum_{n=0}^{N_0}{\hat\epsilon\over\hat\nu_n}
\sum_{m=0}^{N_0}{\hat\epsilon\over\hat\nu_m}
\sin\left(t\log{1\over\hat\nu_n}\right)
\left[\log{1\over\hat\nu_>}\right]
\sin\left(t\log{1\over\hat\nu_m}\right).
\end{equation}
In order to resolve $\hat\nu_>$ we split the double sum, making use of
the symmetry under $n\leftrightarrow m$ interchange, and obtain
\begin{equation}
\label{eq:k2bls}
\langle0_c|\log{1\over\hat\nu_>}|0_c\rangle
=-{8t\over\pi}\sum_{n=0}^{N_0}{\hat\epsilon\over\hat\nu_n}
S_n(t){d\over dt}\cos\left(t\log{1\over\hat\nu_n}\right),
\end{equation}
where
\begin{equation}
\label{eq:sntsum}
S_n(t)=\sum_{m=0}^n{\hat\epsilon\over\hat\nu_m}
\sin\left(t\log{1\over\hat\nu_m}\right)
={\rm{Im}}\left\{\hat\epsilon^{-it}\Big[\zeta(1+it,{\textstyle{1\over2}})-
\zeta(1+it,n+{\textstyle{3\over2}})\Big]\right\}.
\end{equation}
Inserting (\ref{eq:sntsum}) into (\ref{eq:k2bls}),
we find
\beqa
\label{eq:bigone}
\kern-18pt\langle0_c|\log{1\over\hat\nu_>}|0_c\rangle
&=& {2it\over\pi}\Biggl\{
\Bigl[\hat\epsilon^{-it}\zeta(1+it,{\textstyle{1\over2}})
-\hat\epsilon^{it}\zeta(1-it,{\textstyle{1\over2}})\Bigr] \\ \nonumber
&&\times {\partial\over\partial t}\Bigl[
\hat\epsilon^{-it}\Bigl(\zeta(1+it,{\textstyle{1\over2}})
                       -\zeta(1+it,N_0+{\textstyle{3\over2}})\Bigr)
+\hat\epsilon^{it}\Bigl(\zeta(1-it,{\textstyle{1\over2}})
                       -\zeta(1-it,N_0+{\textstyle{3\over2}})\Bigr)
\Bigr]\kern-16pt\\ \nonumber
&&\kern-4em-\hat\epsilon^{-it}{\partial\over\partial t}\Bigl[
\hat\epsilon^{-it}\Bigl(Z(t^\prime,t;{\textstyle{1\over2}})
                       -Z(t^\prime,t;N_0+{\textstyle{3\over2}})\Bigr)
+\hat\epsilon^{it}\Bigl(Z(t^\prime,-t;{\textstyle{1\over2}})
                       -Z(t^\prime,-t;N_0+{\textstyle{3\over2}})\Bigr)
\Bigr]_{t^\prime=t}\\ \nonumber
&&\kern-4em+\hat\epsilon^{it\hphantom{-}}{\partial\over\partial t}\Bigl[
\hat\epsilon^{-it}\Bigl(Z(-t^\prime,t;{\textstyle{1\over2}})
                       -Z(-t^\prime,t;N_0+{\textstyle{3\over2}})\Bigr)
+\hat\epsilon^{it}\Bigl(Z(-t^\prime,-t;{\textstyle{1\over2}})
                       -Z(-t^\prime,-t;N_0+{\textstyle{3\over2}})\Bigr)
\Bigr]_{t^\prime=t}\Bigg\},
\kern-32pt
\eeqa
where,
\beqa
Z(t^\prime,t;a) &=& \sum_{n=0}^\infty\frac{1}{(n+a)^{1+it}}\zeta
(1+it^\prime,1+a+n) \\ \nonumber
 &=& {1\over\Gamma(1+it^\prime)
\Gamma(1+it)}\int_0^\infty dx\int_0^\infty dy{x^{it^\prime}
y^{it}e^{-x}e^{-a(x+y)}\over (1-e^{-x})(1-e^{-x-y})}.
\eeqa
Changing the integration variables to $u=1-e^{-x}$ and $v=1-e^{-x-y}$, we
find for $t\to 0$ and $t^\prime\to 0$ that
\beq
Z(t^\prime,t;a)=-\frac{1}{t^\prime(t+t^\prime)}\frac{\Gamma(a)}
{\Gamma(a+i(t+t^\prime))}+{\cal O}(1).
\eeq
Therefore
\begin{equation}
Z(t^\prime,t;{1\over 2})\simeq-{1\over t^\prime(t+t^\prime)}
+{i\over t^\prime}\psi({1\over 2})
\end{equation}
and
\begin{equation}
Z(t^\prime,t;N_0+{3\over 2})\simeq-{1\over t^\prime(t+t^\prime)}
(N_0+{3\over 2})^{-i(t^\prime+t)}.
\end{equation}
Upon substituting the above expressions into (\ref{eq:bigone}), and
expanding for $t\ll1$, after some manipulations we obtain
\begin{equation}
\langle0_c|\log{1\over\hat\nu_>}|0_c\rangle
\simeq {1\over t^2}[1-{8t\over\pi}(\psi({\textstyle{1\over2}})+\log2)]
={4\over\pi^2}\log^2{2\over\hat\epsilon}[1+4(\gamma+\log2)
\log^{-1}{2\over\hat\epsilon}].
\end{equation}

Finally, after combining this with the discrete normalization,
(\ref{eq:disnorm}), and subtracting out the zeroth order eigenvalue, we
find the claimed result,
\begin{equation}
\langle\tilde0|{\cal K}_1^{(b)}|\tilde0\rangle={4\over\pi^2}
\Bigl[2(\gamma+\log2)\log{2\over\hat\epsilon} + {\cal O}(1)\Bigr].
\end{equation}

In addition, we have also calculated the general inner products
$\langle j_c^{\vphantom{\prime}}|j_c^{\prime}\rangle$ and the matrix element
$\langle0_c|\log\frac{1}{\hat\nu_>}|j_c\rangle$, and found that
\begin{eqnarray}
\langle j_c|j_c\rangle&=&\langle0_c|0_c\rangle,\\
\noalign{\vskip2mm}
\langle j_c^{\vphantom{\prime}}|j_c^\prime\rangle&=&{\cal O}
\left(\log^{-1}\frac{2}{\hat\epsilon}
\right)\qquad\hbox{for $j\neq j^\prime$,}
\end{eqnarray}
and
\beq
\langle0_c|\log\frac{1}{\hat\nu_>}|j_c\rangle=
{\cal O}\left(\log\frac{2}{\hat\epsilon}\right)
\eeq
for $j\neq 0$.

\section{The Discrete Kernel in the Continuum Limit}

Here we demonstrate that in the continuum limit ${\mathcal K}_0$ indeed
leads to the continuum approximate form (\ref{sonsapprox}).
To leading order, there is no distinction between $|j\rangle$,
$j_c\rangle$ and
$|\tilde{j}\rangle$, nor between the eigenvalues of the continuous or
discrete theories.  Thus we are interested in the quantity,
\beq
\label{eq:akap}
\langle \nu_n |{\cal K}_0| \nu_{n'} \rangle = \sum_j
{\widetilde\lambda_j^{\circ\,^{-2}}}
\langle \nu_n |\tilde{j} \rangle \langle \tilde{j} | \nu_{n'} \rangle
\simeq \sum_j {\widetilde\lambda_j^{\circ\,^{-2}}} \langle \nu |j \rangle
\langle j | \nu' \rangle ,
\eeq
where the last expression on the right hand side contains continuous
energy variables and the last inequality holds at leading order only.
Since $\langle \nu | j \rangle = \chi(\nu)$,
using equation (\ref{sol}), we may re-write (\ref{eq:akap}) as,
\beq
\label{intermsI}
\sum_j {\widetilde\lambda_j^{\circ\,^{-2}}}\langle\nu|j\rangle
\langle j|\nu' \rangle = \frac{1}{\pi^2} \log \frac{2}{\hat\epsilon}
\left[I(\theta_\nu - \theta_{\nu'}) - I(\theta_\nu + \theta_{\nu'}) \right],
\eeq
where
\beq
\label{eq:Iafunc}
I(\alpha) \equiv \sum_{j=0}^{\infty} \frac{1}{\left(j + \frac{1}{2} \right)^2}
\cos[(j + {\textstyle\frac{1}{2}})\alpha],
\qquad -\pi \leq \alpha \leq \pi,
\eeq
and
\beq \theta_\nu =
\pi \frac{\log 1/\hat\nu}{\log 2/\hat\epsilon},
\qquad 0 < \theta_\nu \leq \pi.
\eeq
To evaluate $I(\alpha)$ we consider its derivative,
\beq
\label{Ia}
\frac{dI}{d\alpha} = -{\rm Im} \sum_{j=0}^{\infty}
\frac{1}{j + \frac{1}{2}} e^{i(j+\frac{1}{2}) \alpha}
= -{\rm Im}\log \left[ \frac{1 + e^{i \alpha/2}}{1 - e^{i \alpha/2}}\right]
= 2m \pi - \frac{\pi}{2},
\eeq
where $m$ is an integer, yet to be determined.  The condition
$\left. \frac{dI}{d\alpha} \right|_{\alpha = \pi} = -\frac{\pi}{2}$
then fixes $m$ to be zero and $I(\alpha = \pi)=0$ fixes the constant of
integration to give $I(\alpha)$ from (\ref{Ia}).  Since $I(\alpha)$ is an
even and antiperiodic function, we find,
\beq
\label{I}
I(\alpha) = \frac{\pi}{2} \left( \pi - |\alpha| \right), \qquad
-2\pi \leq \alpha \leq 2 \pi.
\eeq
Substituting (\ref{I}) into (\ref{intermsI}) then yields,
\begin{eqnarray}
\langle \nu |{\cal K}_0| \nu' \rangle  =
\cases{\log 1/\hat\nu' & \mbox{ if $\hat\nu'>\hat\nu$,}\cr
\noalign{\vskip2mm}
       \log 1/\hat\nu &  \mbox{ if $\hat\nu'<\hat\nu$,}\cr
}
\label{an4x}
\end{eqnarray}
which is in direct agreement with the leading order kernel written in
(\ref{sonsapprox}).
%

\section{First Order Corrections From ${\mathcal K}_1^{(a)}$.}

In this appendix, we calculate the first order perturbation from
${\cal K}_1^{(a)}$. On splitting the discrete energy sums resulting from
evaluation of the matrix element into diagonal and off-diagonal parts,
\beq
\label{A1}
\langle0|{\cal K}_1^{(a)}|0\rangle=I+J,
\eeq
we have
\beq
\label{A2}
I=\frac{1}{\beta^2}\sum_{n=0}^{N_0}\int_0^\infty dp\int_0^\infty dp^\prime
 {\cal K}_1^{(a)}(n,p|n,p^\prime)\langle\tilde 0|n,p\rangle
\langle n,p^\prime|\tilde 0 \rangle
\eeq
and
\beq
\label{A3}
J=\frac{1}{\beta^2}\sum_{n=0}^{N_0}\sum_{n^\prime=0}^{N_0}
(1-\delta_{nn^\prime})
\int_0^\infty dp\int_0^\infty dp^\prime{\mathcal K}_1^{(a)}(n,p|n',p^\prime)
\langle\tilde 0|n,p\rangle\langle n^\prime,p^\prime|\tilde 0\rangle,
\eeq
where $N_0\gg1$, the discrete cutoff, is given by $\delta=
(N_0+\frac{1}{2})\epsilon$ with
$\delta$ the ultraviolet cutoff introduced in section IV.

For the first term, using (\ref{eq:neqnp}), we find that
\beq
\label{A4}
I=\frac{8t}{\pi\beta^2}\sum_{n=0}^{N_0}\int_0^\infty dp\int_0^\infty dp^\prime
\frac{3\log\frac{p+p^\prime}{|p-p^\prime|}-2\log\frac{1}{\hat\nu_n}}
{\sqrt{[\nu_n^2+(p-\mu)^2][\nu_n^2+(p^\prime-\mu)^2]}}\sin^2\Big(t
\log\frac{1}{\hat\nu_n}\Big),
\eeq
with $t=\frac{\pi}{2\log\frac{2}{\hat\epsilon}}$. For $\epsilon\ll\mu$,
(\ref{A4}) can be approximated by
\begin{eqnarray}
\label{A5}
I&\simeq&\frac{2t}{\pi^3}\sum_{n=0}^\infty\frac{1}{(n+\frac{1}{2})^2}
\sin^2\Big(t\log\frac{1}{\hat\nu_n}\Big)
\int_{-\infty} ^\infty dx\int_{-\infty}^\infty dy\frac{1}{(x^2+1)(y^2+1)}
\nonumber\\
&&\kern6cm\times\Big[3\log\frac{2\mu}{|x-y|(n+\frac{1}{2})\epsilon}
-2\log\frac{1}{(n+\frac{1}{2})\hat\epsilon}\Big]\nonumber\\
&\simeq&\frac{2t}{\pi}\log\frac{2\mu}{\epsilon}\sum_{n=0}^\infty
\frac{1}{(n+\frac{1}{2})^2}\sin^2\Big(t\log\frac{1}{\hat\nu_n}\Big)\nonumber\\
&=&{\cal O}(1)
\end{eqnarray}
for $t\to 0$, since $\log\frac{2\mu}{\epsilon}\sim t^{-1}$ and the
infinite series converges.

Next, instead of calculating $J$ directly, we evaluate the more general
expression,
\beq
\label{A33}
J_m=\frac{1}{\beta^2}\sum_{n=0}^{N_0}\sum_{n^\prime=0}^{N_0}
(1-\delta_{nn^\prime})
\int_0^\infty dp\int_0^\infty dp^\prime {\mathcal K}_1^{(a)}
(n,p|n',p^\prime)
\langle\tilde m|n,p\rangle\langle n^\prime,p^\prime|\tilde m\rangle,
\eeq
so that $J=J_{m=0}$. It follows from (\ref{eq:k1a}) for $\epsilon\ll\mu$,
that
\beqa
\label{A6}
J_m & = & \frac{2t}{\pi}\sum_{n=0}^{N_0}\sum_{n^\prime=0}^{N_0}
(1-\delta_{nn^\prime})\frac{1}{(n+\frac{1}{2})
(n^\prime+\frac{1}{2})}\Big(\log\frac{1}{|\hat\nu_n-\hat\nu_{n^\prime}|}
+\log\frac{1}{|\hat\nu_n+\hat\nu_{n^\prime}|}-2\log\frac{1}{\hat\nu_>}
\Big) \nonumber\\
&&\kern6cm\times \sin\Big[(2m+1)t\log\frac{1}{\hat\nu_n}\Big]
\sin\Big[(2m+1)t\log\frac{1}{\hat\nu_{n^\prime}}\Big] \nonumber\\
& = & \frac{4t}{\pi}\sum_{n=1}^{N_0}\sum_{n^\prime\neq 0}^{n-1}
\frac{1}{(n+\frac{1}{2})(n^\prime+\frac{1}{2})}\sum_{l=1}^\infty\frac{1}{l}
\Big(\frac{\hat\nu_{n^\prime}}{\hat\nu_n}\Big)^{2l}
\sin\Big[(2m+1)t\log\frac{1}{\hat\nu_n}\Big]
\sin\Big[(2m+1)t\log\frac{1}{\hat\nu_{n^\prime}}\Big]
\nonumber\\
&\simeq& \frac{4t}{\pi}\sum_{l=1}^\infty\frac{1}{l}
\int_{\frac{\hat\epsilon}{2}}^{\hat\delta}\frac{d\hat\nu}{\hat\nu}
\int_{\frac{\hat\epsilon}{2}}^{\hat\delta}
\frac{d\hat\nu^\prime}{\hat\nu^\prime}
\Big(\frac{\hat\nu^\prime}{\hat\nu}\Big)^{2l}\sin\Big[(2m+1)t\log\frac{1}
{\hat\nu}\Big]\sin\Big[(2m+1)t\log\frac{1}{\hat\nu^\prime}\Big] \nonumber\\
& = & \frac{4t}{\pi}\sum_{l=1}^\infty\frac{1}{l}
\int_{\log\frac{1}{\hat\delta}}^{\log\frac{2}{\hat\epsilon}}dx
\int_{\log\frac{1}{\hat\delta}}^xdx^\prime e^{-2l(x-x^\prime)}
\sin[(2m+1)tx]\sin[(2m+1)tx^\prime]\nonumber\\
& \sim & \frac{4t}{\pi}\sum_{l=1}^\infty\frac{1}{l}
\int_0^{\log\frac{2}{\hat\epsilon}}dx\int_0^xdx^\prime e^{-2l(x-x^\prime)}
\sin[(2m+1)tx]\sin[(2m+1)tx^\prime]\nonumber\\
& \simeq & \frac{1}{2}\sum_{l=1}^\infty\frac{1}
{l^2+(m+\frac{1}{2})^2t^2}=\frac{\pi^2}{12}
\eeqa
as $t\to 0$. Therefore $J={\cal O}(1)$ and no linearly logarithmic terms
result from $\langle0|{\cal K}_1^{(a)}|0\rangle={\cal O}(1)$. On the other
hand, as $m\to\infty$ at a fixed $t$, $J_m\sim 1/m$. Since $J_m$ is
proportional to the first order shift of the $\frac{1}{\lambda_m^2}$
given by (\ref{eq:evals}), the large $m$ behavior of $J_m$ explains the
weak logarithmic singularity of the Fredholm kernel.

\end{appendix}

\end{document}